\documentclass[12pt]{article}
\usepackage{eqsection}
\usepackage{epsf}

\def \bp {\mbox{\boldmath $\partial$}}

\def\beq{\begin{equation}}
\def\eeq{\end{equation}}
\def\bea{\begin{eqnarray}}
\def\eea{\end{eqnarray}}

\def\nn{\nonumber}

\relax
\begin{document}

\begin{titlepage}
\begin{center}
{\large\bf Two-dimensional O($n$) model in a staggered field}\\[.5in]
{\bf Dibyendu Das and Jesper Lykke Jacobsen}\\[.2in]
{\it Laboratoire de Physique Th\'eorique et Mod\`eles Statistiques, \\
     Universit\'e Paris-Sud, B\^atiment 100, F-91405 Orsay, France.}\\[.2in]
E-mails: das@ipno.in2p3.fr, jacobsen@ipno.in2p3.fr
\end{center}

\underline{Abstract.}

Nienhuis' truncated O($n$) model gives rise to a model of self-avoiding loops
on the hexagonal lattice, each loop having a fugacity of $n$. We study such
loops subjected to a particular kind of staggered field $w$, which for $n \to
\infty$ has the geometrical effect of breaking the three-phase coexistence,
linked to the three-colourability of the lattice faces. We show that at $T =
0$, for $w > 1$ the model flows to the ferromagnetic Potts model with $q=n^2$
states, with an associated fragmentation of the target space of the Coulomb
gas.
For $T>0$, there is a competition between $T$ and $w$ which gives rise to
multicritical versions of the dense and dilute loop universality classes. Via
an exact mapping, and numerical results, we establish that the latter two
critical branches coincide with those found earlier in the O($n$) model on the
triangular lattice. Using transfer matrix studies, we have found the
renormalisation group flows in the full phase diagram in the $(T,w)$ plane,
with fixed $n$.
  

Superposing three copies of such hexagonal-lattice loop models with staggered
fields produces a variety of one or three-species fully-packed loop models on
the triangular lattice with certain geometrical constraints, possessing
integer central charges $0 \le c \le 6$. In particular we show that Benjamini
and Schramm's RGB loops have fractal dimension $D_{\rm f}=3/2$.


\end{titlepage}

\newpage

\section{Introduction}

The O($n$) model plays an important role in the realm of two-dimensional
critical behaviour \cite{deGennes,Zinnbook,Mukamel,Nienhuis82}. In its
lattice version, the partition function at temperature $T$ reads
\beq
 Z_{{\rm O}(n)} = \sum_{\{ \vec{S}_i \}} \prod_{\langle ij \rangle}
 \exp \left( \frac{1}{T} \vec{S}_i \cdot \vec{S}_j \right),
\eeq
where the $\vec{S}_i$ are $n$-component spins living on the vertices $i$
of some regular two-dimensional lattice. The set of nearest-neighbour
vertex pairs (lattice edges) is denoted by $\langle ij \rangle$.

{}From the high-temperature expansion, $Z$ is linked to a model of loops, each
loop having the fugacity $n$ \cite{deGennes}. On the other hand, the O($n$)
model is related to standard $\phi^4$ field theory, where the quartic term
encodes the possibility of loop crossings \cite{deGennes,Zinnbook}. In two
dimensions, this model possesses a second-order phase transition only for $|n|
\le 2$. At the critical point, loop crossings are disfavoured, and in
particular the limit $n \to 0$ gives access to the scaling properties of
self-avoiding loops \cite{deGennes}.

A related model can be obtained by truncating the
high-temperature expansion as follows \cite{Mukamel,Nienhuis82}
\beq
 Z_{\rm loop} = \sum_{\{ \vec{S}_i \}} \prod_{\langle ij \rangle}
 \left( 1 + \frac{1}{T} \vec{S}_i \cdot \vec{S}_j \right),
\eeq
the model now being defined on a regular hexagonal
lattice. The advantage of this truncation---and of the particular choice
of the lattice---is that to all orders in the high-temperature expansion the
loops become strictly self-avoiding. More precisely, one has
\beq
 Z_{\rm loop} = T^{-N} \sum_{\cal C} n^{\cal L} T^{\cal V},
 \label{Zloop}
\eeq
where the summation is over configurations ${\cal C}$ of self-avoiding,
mutually-avoiding loops on a hexagonal lattice of size $N$ vertices.
By construction, the number
of loops passing through a given vertex is either zero or one. ${\cal L}$
denotes the number of loops in a given configuration, and ${\cal V}$ the
number of vacancies, i.e., lattice vertices not visited by any loop.

The model defined by $Z_{\rm loop}$ is exactly solvable (in the sense
of the Bethe Ansatz) along the curves
\cite{Nienhuis82,Baxter86}
\beq
 T^2 = 2 \pm \sqrt{2-n},
 \label{Tsolv}
\eeq
for $|n| \le 2$. Furthermore, it exhibits critical behaviour with
algebraically decaying correlation functions. Its critical exponents along the
curve (\ref{Tsolv})${}^+$ are known exactly
\cite{Nienhuis82,Saleur86,Batchelor88} and are believed to reproduce those of
the generic O($n$) model at its critical point $T_{\rm c}$. We shall refer to
this critical phase as that of {\em dilute loops}. The curve
(\ref{Tsolv})${}^-$ is to be understood as a line of renormalisation group
fixed points that, for each $n \in (-2,2)$, attract the whole low-temperature
region $T \in (0,T_{\rm c})$. Also along this curve the critical exponents are
known exactly \cite{Nienhuis82,DupSal87,Batchelor88}. We shall use the epithet
{\em dense loops} to refer to the corresponding critical geometry.

However, the solution of $Z_{\rm loop}$ along (\ref{Tsolv})${}^-$ does {\em
not} describe the generic behaviour of the O($n$) model in the low-temperature
regime \cite{ReadSal01,JRS03}. The reason is that the model defined by $Z_{\rm
loop}$ is unstable towards the inclusion of loop crossings. When included,
these make the model flow to the generic symmetry-broken (Goldstone) phase of
the $\phi^4$ model, which can in turn be described by supersymmetric methods
\cite{ParSou,ReadSal01,JRS03}. Despite of this fact, $Z_{\rm loop}$ defines a
very interesting model of self-avoiding loops and merits to be studied in
detail on its own right.

Note that the sign of $T$ in (\ref{Tsolv}) is immaterial. This reflects the
fact that, with suitable periodic boundary conditions, the number of vacancies
is necessarily even. Thus, $T=0$ constitutes another line of fixed points
\cite{Reshetikhin,Blote94}, henceforth referred to as {\em compact loops}.
This zero-temperature limit is once again exactly solvable
\cite{Baxter70,Batchelor94}, and for $|n| \le 2$ it is critical with exponents
that are known exactly \cite{Batchelor94,Kondev95}. From a geometrical point
of view, this limit is particularly interesting since only the subset ${\cal
C}_0 \subset {\cal C}$ of loop configurations in which the loops are {\em
fully packed} contributes to $Z_{\rm loop}$.

In this paper we shall study the hexagonal-lattice loop model (\ref{Zloop})
in the presence of a particular kind of staggered field%
\footnote{In section~\ref{sec_staggered} we shall give some justification
for the nomenclature ``staggered field'' and discuss its relation to a
similar construction in the Potts model.}
$w$. To define this field, we label the faces of the hexagonal lattice by
integers $k=1,2,3$ in such a way that any two adjacent faces carry different
labels. Further define $E_0$ as the set of lattice edges whose two adjacent
faces carry the labels $1$ and $3$. Given a loop configuration ${\cal C}$, let
${\cal F}$ be the number of edges in $E_0$ that are covered by a loop segment.
The model to be studied is then defined by the partition function
\beq
 Z = \sum_{\cal C} n^{\cal L} T^{\cal V} w^{\cal F}.
 \label{model}
\eeq
Note that taking $w \neq 1$ does not break those of the lattice symmetries
(translations and rotations) that respect the above sub-lattice structure.
Thus, in the continuum limit we can hope to find further conformally invariant
critical points for this model.

Our main objective is to identify the various critical behaviours
of the model (\ref{model}), and to study its phase diagram and renormalisation
group flows with respect to the parameters $n$, $T$, and $w$. To this end
we employ analytical arguments based on a Coulomb gas construction,
exact mappings to related lattice models, and numerical transfer matrix
results. In particular we show that, for $|n| \le 2$, the model generically
exhibits three more critical points than the dilute, dense, and compact
loops invoked above. The first of these is encountered when $w\to\infty$
with $T=0$, and we shall identify it as a $q=n^2$ state critical Potts model.
The remaining two critical points are superpositions of a critical Ising
model (free fermion) and dilute or dense loops, respectively.

The second part of the paper deals with fully-packed loop (FPL) models on the
triangular lattice. The exact solutions of the FPL models on the hexagonal
\cite{Batchelor94,Kondev95} and the square \cite{Kondev98} lattices imply that
these models have different critical exponents. The reason for this
non-universality can be traced back to a difference in the dimensionality of
the target space of the corresponding Coulomb gases. Indeed, while the usual
dense and dilute loop phases are described by the continuum limit of an
interface model with a scalar height, the heights of the FPL model on the
honeycomb (resp.~square) lattice are vectors of dimension two (resp.~three)
\cite{Kondev95,Kondev98}. This motivates the study of FPL models
on other lattices.

However, it is known numerically \cite{Batchelor96} that the standard
FPL model on the triangular lattice%
\footnote{This is also true for a class of decorated lattices that interpolate
between the square and the triangular lattices \cite{Higuchi99}.} is in the
same universality class as dense polymers [i.e., as the model (\ref{Zloop})
along the curve (\ref{Tsolv})${}^-$]. It was pointed out in
Ref.~\cite{Jacobsen99} that this type of flow from fully-packed loops to the
low-temperature phase of the model (\ref{Zloop}) is indeed the generic
scenario.

To avoid this flow, and to get new interesting critical behavior, one
may subject the triangular-lattice loops to further constraints than
just full packing. One such model, known as the RGB model, was introduced
by Benjamini and Schramm \cite{BenjSch} and studied numerically by
Wilson \cite{Wilson02}. In this model, the fully-packed loops are simply
prohibited to take turns through an angle of $\pm \pi/3$ at any vertex;
each loop carries a trivial fugacity
of $n=1$. However, the critical exponents of this model do not appear to
have been determined \cite{Wilson02}.

Here we study several versions of such constrained FPL models on the 
triangular lattice, and we relate their critical exponents to those of
the model (\ref{model}). In particular, we establish that the fractal dimension
of RGB loops is $D_{\rm f}=3/2$.

The layout of the paper is as follows. In section~\ref{sec_model} we introduce
the model (\ref{model}) and relate it to edge (Tate) colourings of the
hexagonal lattice. We also show an exact mapping of our three-parameter model
to a twelve-parameter loop model on the triangular lattice which has been
studied earlier \cite{Nienhuis98}. In section~\ref{sec_TM} we discuss the
structure of the transfer matrix, which we use to obtain both analytical and
numerical results. Section~\ref{sec_staggered} gives some motivation for the
introduction of the staggered field $w$, by comparing it to a similar
construction for the Potts model. In section~\ref{sec_CG} we fix the notation
by reviewing the Coulomb gas formalism for the FPL model at $w=1$
\cite{Kondev95}. We then show that for $w \to \infty$ the continuum limit of
the FPL model coincides with that of a $q=n^2$ state Potts model. The same
approach also reproduces known results for the finite temperature O($n$) model
\cite{Nienhuis82}. In section~\ref{Onwings} we discuss the results for the
fixed points, critical surfaces, and renormalisation group flows in the full
phase diagram in the $(T,w)$ plane, at any fixed $n$. In section~\ref{sec_tri}
we derive exact results for RGB loops on the triangular lattice
\cite{Wilson02}, and for some generalisations thereof. We give our conclusions
in section~\ref{sec_conclusion}.

\section{The model and some of its reformulations}
\label{sec_model}

Our loop model is defined on a hexagonal lattice with a set of special lattice
edges $E_0$. If the faces of the hexagonal lattice are labeled by the
integers $k=1,2,3$ as shown in Fig.~\ref{lattice}, then $E_0$ is the set of
lattice edges between the faces labeled $1$ and $3$. On such a lattice,
configurations ${\cal C}$ of self-avoiding, mutually-avoiding loops
are laid down with weights given by Eq.~(\ref{model}).

In the present section we review various exact transformations relating this
model to other discrete lattice models. These transformations become useful
in later sections when we dress the Coulomb gas of the model (\ref{model})
and examine its critical properties.

\subsection{Three-colouring model}
\label{sec_3C_model}

When $T=0$, the subset ${\cal C}_0$ of configurations ${\cal C}$
which carry non-zero weight in Eq.~(\ref{model}) are those
in which every vertex is visited by a loop. We shall refer to this as a
fully-packed loops (FPL).

Such loops are related to the three-colouring model introduced
by Baxter \cite{Baxter70}. In this model, the edges of the hexagonal
lattice are covered by colors $A$, $B$ and $C$, subject to the constraint
that no two edges of the same colour meet at a vertex. These colouring
configurations can be brought into contact with FPL configurations
as follows \cite{Baxter70,Kondev95}.

Attribute an auxiliary orientation (clockwise or anti-clockwise) to each loop,
and assign to each configuration ${\cal C}'_0$ of oriented loops a weight so
that the Boltzmann factor of the corresponding un-oriented FPL configuration
${\cal C}_0$ is recovered by summing independently over the two possible
orientations of each loop. Parametrising the loop weight $n$ as%
\footnote{We are mainly interested in $|n| \le 2$, in which case $\lambda$
  is real \cite{Kondev95}. For $n>2$, $\lambda$ is purely imaginary
  \cite{Baxter70}.}
\beq 
 n = {\rm e}^{6 i \lambda} + {\rm e}^{-6 i \lambda} = 2 \cos(6 \lambda),
 \label{n_lambda}
\eeq
this is accomplished by assigning a weight ${\rm e}^{6 i \lambda}$
(resp.~${\rm e}^{-6 i \lambda}$) to each clockwise (resp.~anti-clockwise)
oriented loop. To complete the mapping, a three-colouring configuration is
identified bijectively with a oriented FPL configuration ${\cal C}'_0$ by
letting the colour $C$ (resp.~$B$) represent a loop segment oriented from a
vertex on the even sublattice to a vertex on the odd sublattice (resp.~from
odd to even). [We have here defined the even (resp.~odd) sublattice as the
vertices whose adjacent edges are dual to an up-pointing (resp.~down-pointing)
triangle.] The colour $A$ means that there is no loop segment on the
corresponding edge.

\begin{figure}
\begin{center}
\epsfxsize=9.0cm \epsfysize=8.0cm
\epsfbox{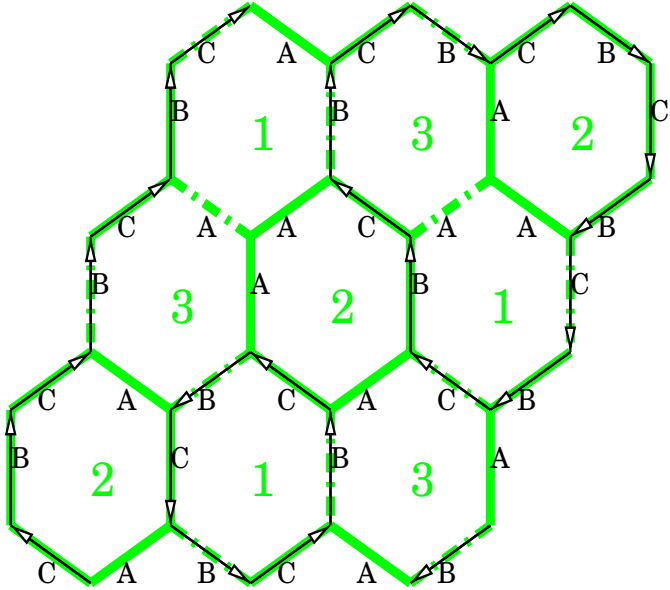}
\caption{Loop configuration on the hexagonal lattice. The lattice faces
are labeled by $1$, $2$ and $3$ as indicated. Among the lattice edges (in
grey), those separating $1$ and $3$-faces (broken linestyle) define the set
$E_0$ (see text). Oriented loops (in black) live on lattice edges with colours
$B$ or $C$. Vertices where three $A$ coloured edges meet break the
fully-packing constraint.}
\label{lattice}
\end{center}
\end{figure}

In this way, oriented loops become alternating sequences $BCBC \cdots$ or
$CBCB \cdots$, depending on the orientation. Also note that violations
of the FPL constraint (with weight $T$) correspond to defects in
which three $A$ colours meet at a vertex. Finally, to recover
Eq.~(\ref{model}), a weight $w$ is attributed to each $E_0$ edge
carrying colour $B$ or $C$. The correspondence between loops
and colours is shown in Fig.~\ref{lattice}.

\begin{figure}
\begin{center}
\epsfxsize=10.0cm \epsfysize=5.0cm
\epsfbox{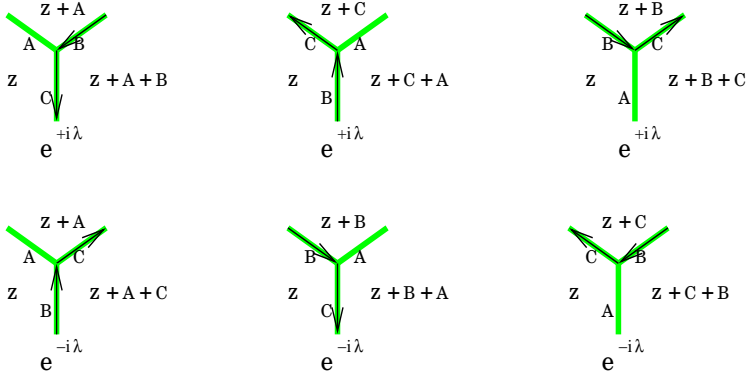}
\caption{The microscopic height increments when crossing the coloured edges,
here shown for an even vertex. We also give
the local vertex weights, corresponding to all possible left and right
turns of directed loop segments (shown as black arrows).}
\label{micro_z}
\end{center}
\end{figure}

The non-local loop weights can now be turned into local vertex weights,
with respect to the three-colouring model, by assigning a weight
${\rm e}^{i \lambda}$ (resp.~${\rm e}^{- i \lambda}$) to each vertex
where an oriented loop turns right (resp.~left).
The weights associated with the different vertex configurations are 
shown in Fig.~\ref{micro_z}. Assuming for the moment free boundary
conditions, every closed loop turns $\pm 6$ times more to the right
than to the left, whence this re-assignment is compatible with
Eq.~(\ref{n_lambda}).

The three-colouring model (with no defects, i.e., $T=0$) can be
mapped to a solid-on-solid model with a two-component microscopic height $z =
(z_1,z_2)$ placed at the vertices of the triangular lattice dual to the
hexagonal lattice. Details of this mapping can be found in
Ref.~\cite{Kondev95},
and are also shown in Fig.~\ref{micro_z}. When going from one vertex of the
dual lattice to its neighbour, moving clockwise around an up-pointing
(resp.~down-pointing) triangle, $z$ is incremented (resp.~decremented) by
the vector ${\bf A}$, ${\bf B}$, ${\bf C}$ depending on the colour of the
direct-lattice edge being crossed. Adopting the normalisation of
Ref.~\cite{Kondev95}, one has
\beq 
 {\bf A} = \left( {1 \over \sqrt{3}}, 0 \right), \qquad
 {\bf B} = \left( -{1 \over 2\sqrt{3}},{1 \over 2} \right), \qquad
 {\bf C} = \left( -{1 \over 2\sqrt{3}}, -{1 \over 2} \right),
\label{def_ABC}
\eeq 
and since ${\bf A}+{\bf B}+{\bf C}={\bf 0}$ the height mapping is well-defined (up to a global shift) for each three-colouring configuration.

It is well-known that in the continuum limit this two-component height
description reduces to a single-component height description when
temperature defects are introduced ($T>0$) \cite{Nienhuis82}.
Below, we shall show that this is also true---in a slightly more subtle
way---when one has $T=0$ and $w>1$.

\subsection{Decimation of $2$-faces}
\label{sec_decimation}

\begin{figure}
\begin{center}
\epsfbox{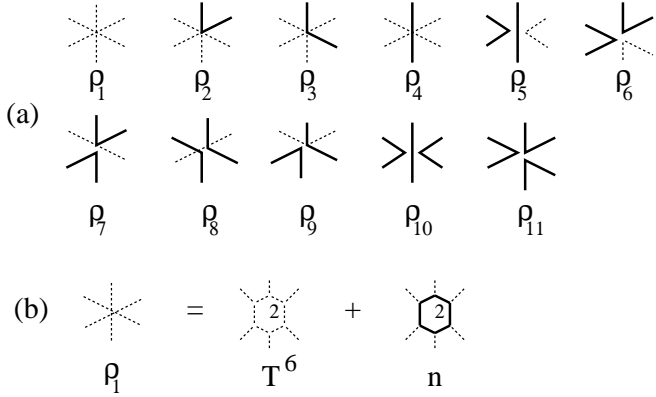}
\caption{(a) Allowed vertices (up to rotations and reflections) of the
triangular-lattice O($n$) model. The vertex weights ${\rho}_{i}$
($i = 1,2,\ldots, 11$) are defined in the figure. (b) As an example, it is
shown how the summation over the two possible configurations around a
$2$-face, given that the six external legs are uncovered, leads to a vertex
on the triangular lattice with weight ${\rho}_{1}=T^6+n$.}
\label{trivertex}
\end{center}
\end{figure}

The model (\ref{model}) can be exactly mapped to a more involved loop model
on the triangular lattice. To see this, consider shrinking the size of all
the $2$-faces in Fig.~\ref{lattice} to zero (alternatively this means shrinking
all edges not in $E_0$ to zero). The result is a loop model on the triangular
lattice in which loops are still non-crossing, but no longer self-avoiding,
in the sense that a given vertex can be visited up to three times by the
loops. Up to rotations and reflections, there are eleven local vertex
configurations, as shown in Fig.~\ref{trivertex}.

The corresponding weights, denoted as $\rho_1,\rho_2,\ldots,\rho_{11}$, can be
related to the three parameters $T$, $w$ and $n$ by summation over the
internal structure (the shrunk $2$-faces) of each triangular-lattice vertex.
Note that all edges remaining after this shrinking are now of the $E_0$ type.
Therefore, the weight $w$ of an occupied edge can be redistributed by
assigning $\sqrt{w}$ to each of the vertices at its extremities. The result is
\bea
 \begin{tabular}{lll}
  $\rho_1 = T^6 + n$ & $\rho_2 = w (T^4 + 1)$  & $\rho_3 = w (T^3 + T)$      \\
  $\rho_4 = 2 w T^2$ & $\rho_5 = \rho_8 = w^2$ & $\rho_6 = \rho_7 = w^2 T^2$ \\
  $\rho_9 = w^2 T$ & $\rho_{10} = 0$         & $\rho_{11} = w^3$           \\
 \end{tabular}
\label{hex_tri_map}
\eea
where, as usual, each closed loop carries a weight $n$.

The model defined by Fig.~(\ref{trivertex}) is the triangular-lattice O($n$)
model previously studied by Knops et al.~\cite{Nienhuis98}, for general values
of the weights $\rho_k$. The exact parametrisations of eight distinct
branches of critical behaviour were conjectured as follows. First, both
the O($n$) model itself and a certain Potts model on the hexagonal lattice
(the latter having both edge and face interactions) were reformulated
as spin-one 141-vertex models on the triangular lattice. The two 141-vertex
models were then supposed to intersect exactly at the critical branches.
The corresponding critical behaviour was then determined, in part by
numerical simulations on the critical branches, and in part by analogy with
an exact solution of the square-lattice O($n$) model \cite{Nienhuis92}.

It is not a priori obvious whether this classification is
complete. Actually this is not the case: for $T=0$ and $w=1$, the model
(\ref{model}) is known to have a compact loop phase
\cite{Batchelor94,Kondev95}, which does not form part of the eight branches
listed in Ref.~\cite{Nienhuis98}.

The mapping (\ref{hex_tri_map}) shows that the three-parameter phase space of
the model (\ref{model}) is a subset of the twelve-parameter phase space of
the triangular-lattice O($n$) model. In particular, in spite of our
incomplete knowledge of the latter model, it seems likely that some of the
critical branches identified in Ref.~\cite{Nienhuis98} are good candidates for
further critical points of the model (\ref{model}). We shall see below that
these expectations are indeed fulfilled.

It is of course unlikely that a putative new critical point of the model
(\ref{model}) has vertex weights (\ref{hex_tri_map}) that exactly
coincide with one of the critical branches of the triangular-lattice O($n$)
model. However, it may still map to a point which {\em flows} to one of these
branches under the renormalisation group (RG). Needless to say, it would be
a tremendous task to elucidate the RG flows of the triangular-lattice O($n$)
model; for the model (\ref{model}) this is however feasible (due to the vastly
smaller number of parameters) and will be done below.

\section{Transfer matrices}
\label{sec_TM}

Before moving on to our analytical results for the model (\ref{model})
we describe the construction of its transfer matrix. Understanding the
conservation laws of the transfer matrix will be helpful in identifying
the operators that are present in the continuum limit. Furthermore,
diagonalising the transfer matrix numerically will serve as a check---and
sometimes as a guideline---for the analytical results. We shall also
use the numerical determination of the central charge and the critical
exponents as an aid in dressing the complete phase diagram of the model,
in section~\ref{Onwings}.

\subsection{Basis states and algorithmic details}

When dressing the transfer matrix for the model (\ref{model}) a natural first
question is how to deal with the non-local loop weights $n$. The solution to
this problem is to work in a space of basis states containing non-local
information about how the loop segments in a given time-slice were
interconnected at former times \cite{Blote89a}. The transfer matrix ${\cal T}$
time propagates the system by updating this connectedness information, and at
the same time builds up the partition function.

The standard power method provides a convenient way of diagonalising ${\cal
T}$. Namely, when letting ${\cal T}$ act repeatedly on a suitable reference
state, the result will converge to the dominant eigenvector, and the rate of
growth of the norm of the iterated state will yield the largest eigenvalue
(which is non-degenerate by the Perron-Frobenius theorem). The reference state
is typically taken as one of the basis states. Care must be taken, however,
since ${\cal T}$ is often block-diagonal due to the existence of conservation
laws (see below). In this case, one must ensure that the reference state
belongs to the same block as the dominant eigenvector.

The power method also allows for extracting sub-dominant eigenvalues,
by iterating several states which are kept mutually orthogonal after
each iteration.

One can think of the entries of ${\cal T}$ as being indexed by a pair of
integers (row and column number). Relating these integers to the connectedness
information which constitutes the basis states is technically involved
\cite{Blote89a}, and is once again linked to the issue of conservation laws.
Fortunately, this complication can be eliminated completely by characterising
each state by a large integer (which is usually easy) and inserting the states
generated at each iteration into a hash table.

Finally, the efficiency of the transfer matrix algorithm is optimised
by using a sparse matrix factorisation scheme, in which ${\cal T}$ is
written as a product of matrices which each add a single vertex to the
lattice.

\subsection{Conservation laws}

We now examine the conservation laws associated with the transfer
matrices for strips of finite width $L$ of the lattice model (\ref{model}).

On the hexagonal lattice there are two natural transfer (or time-like)
directions, ${\cal T}_\parallel$ and ${\cal T}_\perp$, which are respectively
parallel (${\cal T}_\parallel$) and perpendicular (${\cal T}_\perp$) to one
third of the lattice edges. In Fig.~\ref{lattice}, ${\cal T}_\parallel$ is
vertical and ${\cal T}_\perp$ is horizontal. With periodic transverse boundary
conditions, the minimal number of dangling edges $L$ for these two choices are
respectively $3k$ and $2k$ (with $k$ being an integer) in the ground state
sector. (In other words, as a function of $L$ we can expect the various
eigenvalues to exhibit ``mod 3'' and ``mod 2'' fluctuations respectively. The
ground state sector is such that periodic transverse boundary conditions
respect the sublattice structure of the hexagonal lattice.)

In the FPL model (with $T=0$), for the direction ${\cal T}_\parallel$
the number of edges that are occupied by loop segments (i.e., the number of
$B$ colours plus the number of $C$ colours) is conserved \cite{Baxter70}.
For entropic reasons, we can expect the largest eigenvalue to reside in
the sector with the largest number of configurations, which is then the
one with loop density $2/3$. If there are $2l$ occupied loop segments 
in a strip of size $L = 3k$, then the number of states in the sector is 
\beq
 {3k \choose 2l} c_l,
\eeq
where $c_l = (2l)!/[{l!(l+1)!}]$ are the Catalan numbers. 
For the direction ${\cal T}_\perp$, the conserved
quantity is $Q =$ [the number of loop segments on even edges] minus
[the number of loop segments on odd edges]. This can be easily checked 
by considering the action of the transfer matrix on all
possible states of an even edge and the adjacent odd edge.
For $Q = 2 q$, the number
of states in the sector labeled by $Q$ is
\beq
 \sum_{l=0}^{k-2q} {k \choose l+2q} {k \choose l} c_{l+q}.
\eeq
Namely, out of the available $2k$ edges, one has to choose $l$ occupied odd
edges and $l+2q$ occupied even edges. These can then finally be interconnected
in $c_{l+q}$ ways. The largest eigenvalue is expected to reside in the $Q=0$
sector.

Taking $T>0$, these conservation laws are modified. For instance, in the
${\cal T}_\parallel$ direction it is now the difference of the up and down oriented
vertical loop segments (i.e., the number of $B$ colours minus the number of
$C$ colours) which is conserved.

\subsection{Series expansions}

We further note that in the large-$n$ limit, for $w < 1$ and $w > 1$, the 
first few terms of the series expansion for the free energy per face can be
easily written down: 
\bea
-f_{w<1} &=& {1 \over 3} \ln~n + {{2 w^3} \over {3 n^2}} + O({1 \over n^4}) \nn \\ 
-f_{w>1} &=& {1 \over 3} \ln~n + \ln~w + {1 \over {3 n^2}}(1 + {1 \over w^3} + 
\cdots) + O({1 \over n^4}).
\label{series_nlarge} 
\eea
Comparing these series expansions with our numerically obtained free energies 
(see below) provides a useful check of our algorithms.

\subsection{Eigenvalues and critical exponents}

The central charge $c$ and the scaling dimensions $x_i$ of various 
correlation functions can be extracted from the transfer matrix spectra
in a standard way. Let the eigenvalues be labeled as
$\Lambda_0 > \Lambda_1 \ge \Lambda_2 \ge \ldots$ (the first few dominant
eigenvalues turn out to be real and positive).
One can obtain $c$ and the $x_i$ from \cite{Blote86,Affleck86}
\beq
f_0(L) = f(\infty) - {{\pi c} \over {6 L^2}} + o\left(L^{-2}\right)
\label{finite_c}
\eeq
and \cite{Cardy84}
\beq 
f_i(L) - f_0(L) = {{2 \pi x_i} \over L^2} + o\left(L^{-2}\right). 
\label{finite_x}   
\eeq 
Here the free energy per unit area reads $f_0 = -A\ln(\Lambda_0)/(L M)$, where
$M \gg 1$ is the number of rows of the strip; we similarly relate $f_i$ to
$\Lambda_i$. The geometrical factor $A$ reads $A = 2/\sqrt{3}$ for the ${\cal
T}_{\parallel}$ direction and $A = \sqrt{3}/2$ for the ${\cal T}_{\perp}$
direction; it ensures that the free energy is properly normalised per unit
area.

\subsection{The case $T=0$, $w\to\infty$}
\label{sec_TM_w_inf}

\begin{figure}
\begin{center}
\epsfbox{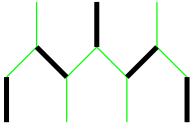}
\caption{A small portion of the hexagonal lattice between two time slices;
the transfer direction is vertical. The $E_0$ edges are shown in black.}
\label{TM_w_inf}
\end{center}
\end{figure}

Consider now the transfer direction ${\cal T}_\parallel$ in the case of $T=0$
and $w\to\infty$. In Fig.~\ref{TM_w_inf} we show a small portion of the
lattice between two time slices; the transfer direction is here
understood to be vertical. The $E_0$ edges are marked on the figure, and
in any allowed configuration these must be covered by loop segments.

We now focus on the two non-$E_0$ vertical edges in the bottom of
Fig.~\ref{TM_w_inf}. Using the fully-packing constraint---and the fact
that all $E_0$ edges must be occupied---it is easy to see that if
these two edges are both unoccupied one arrives at a contradiction.
The same is true if they are both occupied.
Therefore, exactly one of them must be occupied. This implies that
the only allowed sector of the transfer matrix is the one in
which {\em exactly} $2/3$ of the vertical edges are covered.

The ground state sector has $3k$ vertical edges in each time slice.
The number of covered vertical edges is therefore $2k$, i.e., even.

\section{The staggered field}
\label{sec_staggered}

In the introduction, it was claimed that the parameter $w$ in Eq.~(\ref{model})
can be considered as a staggered field. To make this more clear, we wish to
compare it with a similar construction in the Potts model.

\begin{figure}
\begin{center}
\epsfxsize=6.0cm \epsfysize=4.5cm
\epsfbox{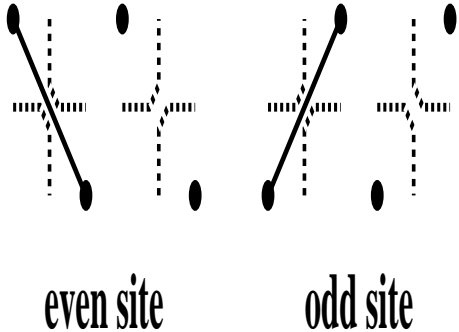}
\caption{The Potts model as a loop model on the square lattice. The circles
represent the Potts spins, the solid lines are bonds in the clusters, and
the broken lines are loop segments.}
\label{fig_FK}
\end{center}
\end{figure}

One of the nicest reformulations of the square-lattice $q$-state Potts model
is as a loop model on the square lattice \cite{Baxter82}. It is well-known
that the Potts model can be transformed into a random cluster model, where the
summation over Potts spins is turned into a summation over
bond-percolation clusters \cite{FK}; each cluster connects vertices whose
spins are in the same state. This can be further transformed into a loop model
in which the loops bounce off the boundaries (both exterior and interior) of
the clusters, and of the dual clusters. This
construction is illustrated in Fig.~\ref{fig_FK}.

The partition function is then proportional to
\beq
 Z_{\rm Potts} = \sum_{\rm loops} q^{{\cal L}/2} u^{\cal B},
\eeq
where ${\cal L}$ is the number of closed loops and ${\cal B}$ is the number
of edges covered by a bond of the clusters. The parameter
$u=({\rm e}^K-1)/\sqrt{q}$, where $K$ is the reduced spin-spin coupling in
the Potts model. According to Fig.~\ref{fig_FK}, the correspondence between
the weight $u$ and the local behaviour of the loops at the clusters
depends on the sublattice. In this sense, $u$ is a staggered field. Also
note that the selfdual point is at $u=1$, and only in this case is the 
model exactly solvable \cite{Baxter82}.

Now consider the limit $q\to\infty$, with $u=1$. There are just two dominant
configurations, each with the same weight, in which loops of length four
encircle the even (resp.~the odd) faces of the lattice. In terms of the spins
these are the completely ordered (ferromagnetic) and the completely disordered
(paramagnetic) states. Moving $u$ away from 1 will favour one of these
configurations. Thus, the selfdual manifold $u=1$ can be considered as a
two-phase coexistence curve, even for finite $q$. For $q>4$ this phase
coexistence is first order: taking $u$ through 1 drives the system through a
first order thermal phase transition with non-vanishing latent heat
\cite{Baxter73}. For $q \le 4$ the coexistence becomes second order, and $u=1$
is indeed the (ferromagnetic) critical point of the Potts model. Even in this
latter case, taking $u \neq 1$ will favour one of the phases and induce an RG
flow that will take the system to either of the two reference configurations,
i.e., to non-criticality.

Our motivation for introducing the model (\ref{model}) is taken from the
above scenario. The parameter $w$ is a staggered field in the above sense.
Consider first the case of $T=0$.
When $n\to\infty$ with $w=1$, there is a coexistence between {\em three}
dominant configurations, in which loops of length six encircle the faces
labeled $k=1,2,3$ respectively.

\begin{figure}
\begin{center}
\epsfxsize=12.0cm \epsfysize=9.0cm
\epsfbox{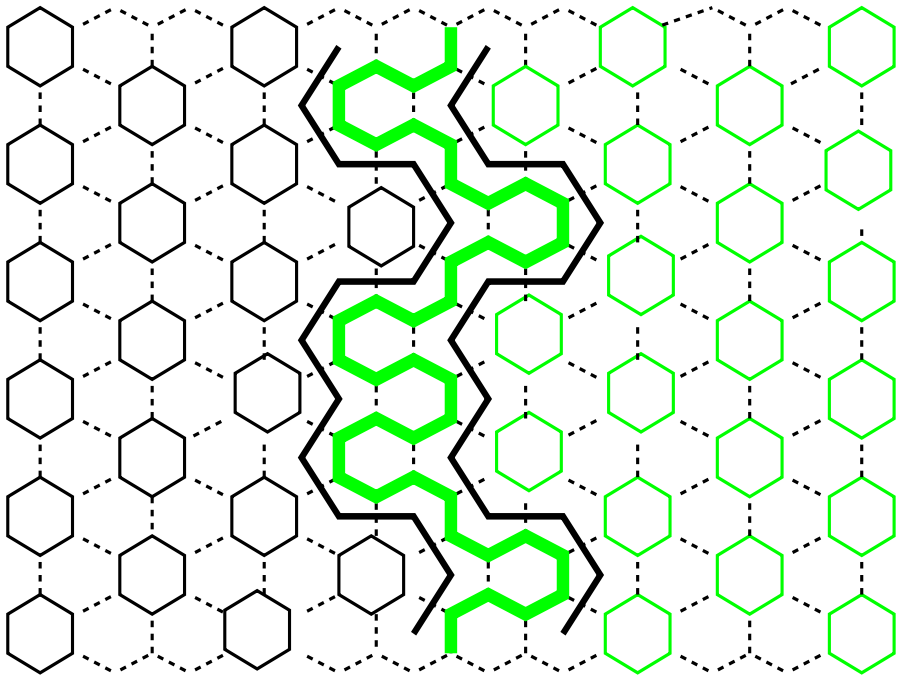}
\caption{The space between the two black curves is filled up by a domain
wall (thick grey loop segment) that separates two domains of
phase 1 (black loops, on the left) and phase 3 (grey loops, on the right).
Potts spins lie at the centres of the hexagons, i.e., on two triangular
lattices, which have a slight mismatch.}
\label{loop_potts}
\end{center}
\end{figure}

In Fig.~\ref{loop_potts} we illustrate a domain wall between faces
$k=1$ and $k=3$, for $n \gg 1$. The states on either side of the domain wall
can be thought of as completely ordered phases of a three-state Potts model,
with spins in the state $k$ being placed at the centres of the small loops.
Let us write its hamiltonian as follows:
\beq
 {\cal H}_{\rm 3-Potts} =
 - K \sum_{\langle ij \rangle} \delta_{\sigma_i,\sigma_j}
 - H \sum_i \left(1 - \delta_{\sigma_i,2} \right).
\label{3-Potts}
\eeq
The Potts spins $\sigma_i=1,2,3$ thus live on a triangular lattice, which is
however slightly shifted across the domain wall. To ensure full packing, a
single long loop is necessary to separate the phases. The space taken up by the
domain wall is delimited by the two black curves. In the ordered phases,
per Potts spin there is one loop of weight $n$ and three satisfied couplings
$K$. Therefore%
\footnote{Note that when $q\to\infty$, the loop model based on the
$q$-state Potts model can be similarly related to an Ising model.
The simple argument above then leads to the relation
$K_{\rm Ising} = \frac18 \log q$; this was found in Ref.~\cite{Cardy97}
by a more complicated argument, which involved
counting the length of the domain wall.}
\beq
 K \sim \frac13 \log n.
\eeq
Taking $w \neq 1$ similarly induces a magnetic field $H$ in the
three-state Potts model (\ref{3-Potts}), with
\beq
 H \sim 3 \log w.
\eeq

\begin{figure}
\begin{center}
\epsfxsize=14.0cm \epsfysize=6.0cm
\epsfbox{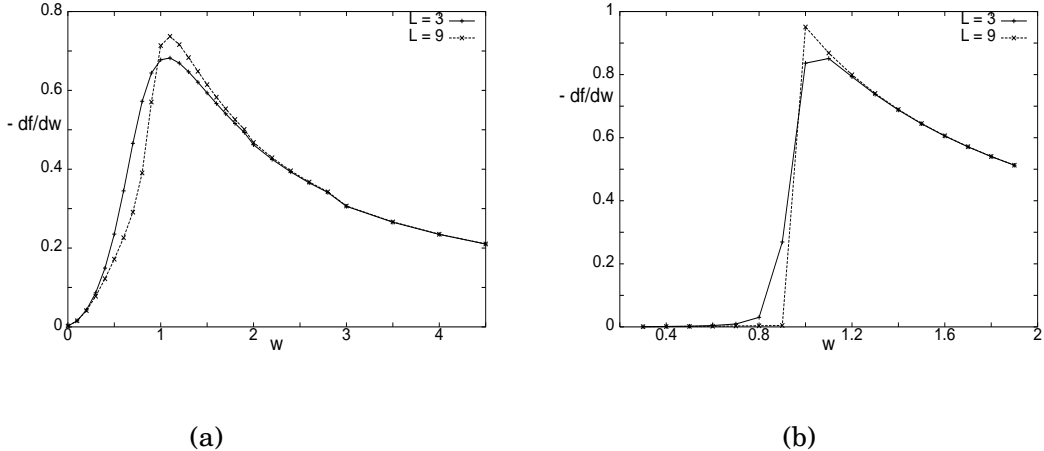}
\caption{$(a)$ The derivative $-df/dw$ at $w=1$ is continuous for $n = \sqrt{3}$. (b) The derivative is discontinuous at $w=1$ for $n = 20$.}
\label{neg_dfree}
\end{center}
\end{figure}

For $n$ finite, the curve $w=1$ can still be considered as the three-phase
coexistence curve. Since the correlation length for $w=1$ is known to be
finite for $n>2$ and infinite for $n \le 2$
\cite{Baxter70,Batchelor94,Kondev95}, it seems natural to expect that taking
$w$ through 1 will induce a first-order transition for $n>2$ and a
second-order transition for $n \le 2$. This expectation is confirmed in
Fig.~\ref{neg_dfree} by numerically evaluating the derivative of the free
energy density $f$ with respect to $w$, using the transfer matrix
${\cal T}_\parallel$ for strips of various widths $L$.

The staggered field $w$ can be expected to have a richer behaviour than
in the Potts case. Namely, for $w>1$ the three-phase coexistence is broken
down to two-phase coexistence (rather than just to a trivial one-phase
``coexistence'', which happens for $w<1$). This means that it is still
possible to find some critical behaviour for $w>1$; in section~\ref{sec_CG}
we shall describe exactly how this comes about.

Consider finally the case of finite temperature $T>0$. The parameter $T$
acts so as to avoid full packing, while $w$ tries to enforce it. There
is thus some hope that the competition between $w$ and $T$ may lead to
interesting multicritical behaviour. This gives rise to a rather rich
phase diagram, which will be investigated in section~\ref{Onwings}.

\section{Coulomb gas formalism}
\label{sec_CG}

The critical phases of loop models can be described in terms of effective
field theories. Exploiting the various symmetries at both the microscopic and
coarse-grained levels, one can construct such theories, and obtain the central
charge and critical exponents exactly via the Coulomb gas correspondence
\cite{Nienhuis87,Kondev95,Kondev98}. 

In section \ref{FPLw1} we begin by reviewing the case of compact loops ($T=0$
and $w=1$), following Refs.~\cite{Kondev95,Kondev98}. Apart from fixing the
notation, this will also serve as a basis for section \ref{FPLw_neq1}, in
which we obtain the central charge and the critical exponents for $w > 1$. In
section \ref{O_n}, we show how the same formalism gives back the known results
\cite{Nienhuis82} for the case of $T>0$ and $w=1$.

\subsection{FPL model at $w = 1$}
\label{FPLw1} 

The effective field theory for the long-wavelength behaviour of the model is
written in terms of a coarse-grained height field, obtained from the
microscopic height $z$ already defined above.

A typical configuration of the critical oriented FPL model, considered in
terms of the colouring model, has domains of {\em ideal states}, which are
colouring states having the least possible variance of $z$. In other word,
ideal states are macroscopically flat. In an ideal state, all loops have
length six and the same orientation: as each loop can be oriented in two ways,
there are six such states. To each of them we assign a coarse-grained height
${\bf h} = (h_1,h_2) = \langle z \rangle$, which is the average microscopic
height over a $\sqrt{3} \times \sqrt{3}$ unit cell of the colouring.

The dominant contributions to the free energy are bounded fluctuations around
these flat states, as they maximise the local entropy density. In the continuum
limit, the height is assumed to be a smooth function of the $2-$dimensional
coordinates $(x^1,x^2)$. The part of the partition function accounting for
the large scale fluctuations, namely $Z_{>} = \int {\cal D} {\bf h} \exp(-
S[{\bf h}])$, has an Euclidean action $S$ with three terms:
\beq 
 S = S_{\rm E} + S_{\rm B} + S_{\rm L}.
\label{action}  
\eeq 

\begin{figure}
\begin{center}
\epsfxsize=14.0cm \epsfysize=10.0cm
\epsfbox{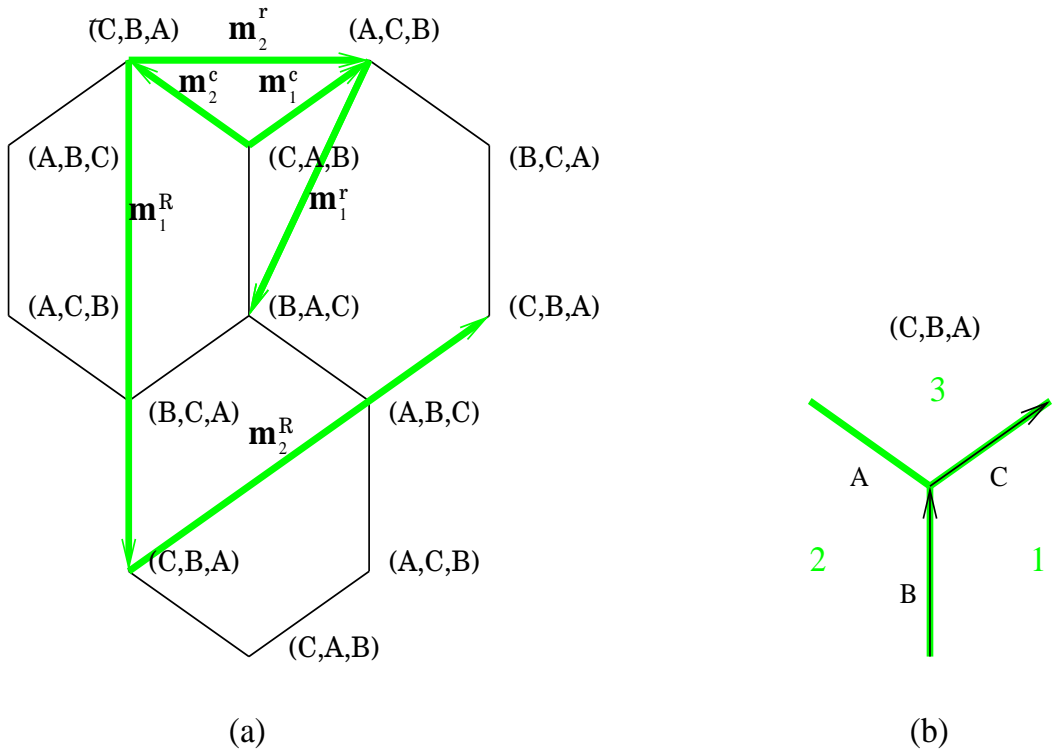}
\caption{$(a)$ The ideal state graph ${\cal I}$ shows the relative positions
of the different ideal states in the coarse-grained height space. $(b)$ The
labeling of an ideal state---here $(C,B,A)$---is the ordered list of
edge colours around an even vertex, with face labels as shown
(starting from the rightmost edge, and turning clockwise around the vertex).
The vectors ${\bf m}^{c}_{1}$, ${\bf m}^{c}_{2}$, ${\bf m}^{r}_{1}$,
${\bf m}^{r}_{2}$, ${\bf m}^{R}_{1}$ and ${\bf m}^{R}_{2}$ are defined in
the text.}
\label{ideal}
\end{center}
\end{figure}

In Fig. \ref{ideal}, we show the positions of the ideal states in the
coarse-grained height
space. The hexagonal lattice with sides ${1 \over 3}$ having the ideal states
at the vertices is referred to as the {\em ideal state graph} ${\cal I}$. The
nodes of ${\cal I}$ that correspond to the same ideal state, form a triangular
lattice with sides $1$, and is referred to as the {\em repeat lattice} ${\cal
R}$.

Below we list some important height differences associated with symmetry
transformations between a pair of ideal states, the order being that of
increasing norm. The vectors ${\bf m}^{c}_{1} = {1 \over 3} ({\bf A} - {\bf
C})$ and ${\bf m}^{c}_{2} = {1 \over 3} ({\bf B} - {\bf A})$ (the superscript
$c$ indicates `colour') have norm ${1 \over 3}$; they are associated with
elementary colour exchanges $A \leftrightarrow C$ and $A \leftrightarrow B$,
respectively; these vectors span a hexagonal lattice. The vectors ${\bf
m}^{r}_{1} = {\bf C}$ and ${\bf m}^{r}_{2} = {\bf A}$ (the superscript $r$
indicates `rotation') have norm ${1 \over \sqrt{3}}$; they are associated with
transformation of one ideal state to another via $120^{\rm o}$ spatial
rotation; they span a triangular lattice. Finally, the vectors ${\bf
m}^{R}_{1} = ({\bf C} - {\bf B})$ and ${\bf m}^{R}_{2} = ({\bf A} - {\bf C})$
(the superscript $R$ stands for `repeat') with norm $1$ form the basis of the
triangular repeat lattice ${\cal R}$; they are associated with transforming
the ideal states back on to themselves.


Each of the terms in Eq.~(\ref{action}) can be interpreted from geometrical
and symmetry considerations.

The first term $S_{\rm E}$ is due to the {\em elastic fluctuations} of the
interface, and has terms in gradient-square of the height field. Imposing
translational and rotational symmetry at the coarse-grained level, the general
form of this term must be $\int {\rm d}^{2}x \, \{K_{11} (\bp h_1)^2 + K_{22}
(\bp h_2)^2 + K_{12} (\bp h_1) \cdot (\bp h_2)\}$, where $\bp = (\partial_1,
\partial_2)$ is the usual gradient. Due to colour exchange symmetries at the
microscopic level, there are some constraints on these coupling constants.
First, the symmetry $B \leftrightarrow C$ changing the loop orientation
corresponds to $z_2 \leftrightarrow -z_2$, $z_1 \leftrightarrow z_1$ in
Eq.~(\ref{def_ABC}) and keeps the weights in Fig.~\ref{micro_z} invariant if
we let $\lambda \leftrightarrow -\lambda$. This implies that $K_{12} = 0$.
Second, the cyclic exchange of colours $A \rightarrow B \rightarrow C$
corresponds to $z_1 \rightarrow -{1 \over 2}z_1 - {\sqrt{3} \over 2}z_2$, $z_2
\rightarrow {\sqrt{3} \over 2}z_1 - {1 \over 2}z_2$ and leaves the weights
unchanged. This implies $K_{11} = K_{22}$; we shall denote this common value
by $g \pi$ henceforth.%
\footnote{We have here sticked to the notation of Ref.~\cite{Kondev95}.
  To compare our formulae with Ref.~\cite{Kondev98}, one needs to replace $g$
  by $K/{2\pi}$ and ${\bf e}$ by ${\bf e}/{2 \pi}$.}
Thus, 
\beq 
 S_{\rm E} = \int {\rm d}^{2}{\bf x} \, \exp \left( {g \pi} \{({\bp} h_1)^2 + ({\bp} h_2)^2\} \right).
\label{elastic}
\eeq

The mapping of the loop model to an oriented loop model with local complex
weights as defined in section \ref{sec_model} does not work for loops which
experience the boundary. For example if the model is defined on a cylinder,
an oriented loop winding around the cylinder will pick up a total weight of
{\em one} since it makes an equal number of left and right turns. Summing
over the two orientations there is a weight of $2$ for the loop, whereas,
by definition of the model, every loop should have an weight $n$. To correct
for this, one introduces a {\em background term} $S_{\rm B}$ in the action:
\beq
 S_{\rm B} = {i \over 2} \int d^2{\bf x} ({\bf e}_0 \cdot {\bf h})
 {\tilde {\bf {\cal R}}}.
\label{background}
\eeq 
In the above, ${\tilde {\bf {\cal R}}}$ is the scalar curvature of the
space on which the model is defined, and ${\bf e}_0$ is the {\em background
electric charge}, yet to be determined.

On the cylinder, the scalar curvature is zero, except at the two boundaries at
infinity along the cylinder axis, so that $S_{\rm B} =\exp(2 \pi i {\bf e}_0
\cdot ({\bf h}(x^1,+\infty) - {\bf h}(x^1,-\infty)))$; we have here taken 
$x^2$ to denote the time-like coordinate.
The unique solution that attributes the correct phase ${\rm e}^{\pm 6 i
\lambda}$ to every winding oriented loop reads
\beq
 {\bf e}_0 = (0, e_0) = \left( 0, {{6 \lambda} \over \pi} \right)
 = \left( 0, {1 \over \pi} \cos^{-1} \left( n/2 \right) \right).
\label{e_o}
\eeq

The terms in the action described so far constitute a Coulomb gas
\cite{CFTbook} with two bosonic fields $h_1$ and $h_2$, where $h_2$ is
coupled to a background charge $-2 e_0$. A crucial role is played by
the last {\em Liouville term} in the action, which reads
\beq
 S_{\rm L} = \int {\rm d}^2{\bf x} \, W[{\bf h}({\bf x})],
\label{liouville}
\eeq
where $\exp(- W[{\bf h}({\bf x})])$ is the scaling limit of the complex 
weights at any vertex, namely $\exp(\pm i \lambda)$ as discussed before. We 
list below the microscopic weights $W$ with respect to colorings 
around the vertex using the notation of Fig.~\ref{ideal}$(b)$
and Eq.~(\ref{e_o}):
\begin{eqnarray}
W(C,A,B) = W(B,C,A) = W(A,B,C) = +i {\pi \over 6} e_0 \nonumber \\
W(C,B,A) = W(B,A,C) = W(A,C,B) = -i {\pi \over 6} e_0
\label{w_micro}   
\end{eqnarray}  
The microscopic operator $W({\bf x})$ is {\it uniform} in each ideal state, 
and is a function of the height ${\bf h} \in {\cal I}$. Since it is a periodic
function of ${\bf h({\bf x})}$ it can be written as a Fourier series,
\beq
W[{\bf h({\bf x})}] = \sum_{{\bf e} \in {\cal R}_{W}^{*}} {\tilde W}_{\bf e}
\exp(i 2 \pi {\bf e} \cdot {\bf h({\bf x})}) .
\label{w_coarse}
\eeq  
In the continuum limit, this becomes a sum of vertex operators, and ${\bf e}$
are the corresponding electric charges which belong to a lattice ${\cal
R}_{W}^{*} \subset {\cal R}^{*}$. Here, ${\cal R}^{*}$ is the reciprocal of
the repeat lattice ${\cal R}$, and ${\cal R}_{w}^{*}$ is the lattice
reciprocal to the lattice of {\em periods} of $W({\bf h})$. We now determine
which vertex operators in Eq.~(\ref{w_coarse}) are the most relevant; these
are the only ones to be kept in the action.

In a Coulomb gas, operators are associated with electric and magnetic charges.
Electric charges ${\bf e}$ are linked to the periodicity of the height field,
and appear in the vertex operators (spins waves)
$\exp(i 2\pi {\bf e} \cdot {\bf h})$, as discussed in Eq.~\ref{w_coarse}.
Magnetic charges ${\bf m}$ are the topological charges of vortex defects
in the height. Within the Coulomb gas formalism, the scaling dimension of
a general operator with electric and magnetic charges ${\bf e}$ and ${\bf m}$
is computed as \cite{Dotsenko84}
\beq 
 x({\bf e},{\bf m}) = {1 \over {2 g}}
 {\bf e} \cdot ({\bf e} - 2 {\bf e}_0) + {g \over 2} {\bf m} \cdot {\bf m}.
\label{dimension}
\eeq

The lattice ${\cal R}$ is spanned by the vectors ${\bf m}^{R}_1 = (0,-1)$ and
${\bf m}^{R}_2 = ({\sqrt{3} \over 2},{1 \over 2})$. The corresponding
reciprocal lattice ${\cal R}^{*}$ is hexagonal, and we denote its six shortest
vectors by ${\bf e}^R$. For future reference, we list here these vectors as
well as the scaling dimensions (found from Eq.~(\ref{dimension})
with ${\bf m} = {\bf 0}$) of the corresponding vertex operators:
\beq
\begin{tabular}{llll}
 ${\bf e}^{R}_{1} = {2 \over \sqrt{3}} \left( {1 \over 2},
                    {-\sqrt{3} \over 2} \right)$,  &
 ${\bf e}^{R}_{4} = {2 \over \sqrt{3}} \left( {-1 \over 2},
                    {-\sqrt{3} \over 2} \right)$   & \qquad \qquad &
 $x({\bf e}^{R}_{1}) = x({\bf e}^{R}_{4}) = 
 {{{4 \over 3} + 2 e_0} \over {2 g}}$ \\
 ${\bf e}^{R}_{2} = {2 \over \sqrt{3}} (1,0)$,     &
 ${\bf e}^{R}_{5} = {2 \over \sqrt{3}} (-1,0)$     & &
 $x({\bf e}^{R}_{2}) = x({\bf e}^{R}_{5}) =
 {{{4 \over 3}} \over {2 g}}$ \\
 ${\bf e}^{R}_{3} = {2 \over \sqrt{3}} \left( {1 \over 2},
                    {\sqrt{3} \over 2} \right)$,   &
 ${\bf e}^{R}_{6} = {2 \over \sqrt{3}} \left( {-1 \over 2},
                    {\sqrt{3} \over 2} \right)$    & &
 $x({\bf e}^{R}_{3}) = x({\bf e}^{R}_{6}) =
 {{{4 \over 3} - 2 e_0} \over {2 g}}$ \\
\end{tabular}
\label{repeat_reci}
\eeq

To find the electric vectors appearing in the expansion (\ref{w_coarse}),
we note that the loop-weight operator $w({\bf h})$ has a higher periodicity
than that of ${\bf m}^{R}$. Namely, the microscopic weights (\ref{w_micro})
of the ideal states are invariant with with respect to $120^{\rm o}$
rotations, which are linked to the periodicity of ${\bf m}^{r}$.
Thus, the lattice ${\cal R}_W$ determining the periodicity of the weights
is spanned by the vectors ${\bf m}^{r}_{1}$ and ${\bf m}^{r}_{2}$ of
Fig.~\ref{ideal}. The corresponding reciprocal lattice ${\cal R}^{*}_W$
is hexagonal, and its shortest vectors ${\bf e}^r$ and the scaling dimensions
of the corresponding vertex operators are as follows:
\beq
\begin{tabular}{llll}
 ${\bf e}^{r}_{1} = 2 (0,-1)$ & & \qquad \qquad &
 $x({\bf e}^{r}_{1}) = {{4 + 4 e_0} \over {2 g}}$ \\
 ${\bf e}^{r}_{2} = 2 \left( {\sqrt{3} \over 2},{-1 \over 2} \right)$, &
 ${\bf e}^{r}_{3} = 2 \left( {-\sqrt{3} \over 2},{-1 \over 2} \right)$ & &
 $x({\bf e}^{r}_{2}) = x({\bf e}^{R}_{3}) =
 {{4 + 2 e_0} \over {2 g}}$ \\
 ${\bf e}^{r}_{5} = 2 \left( {-\sqrt{3} \over 2},{1 \over 2} \right)$, &
 ${\bf e}^{r}_{6} = 2 \left( {\sqrt{3} \over 2},{1 \over 2} \right)$ & &
 $x({\bf e}^{r}_{5}) = x({\bf e}^{r}_{6}) = {{4 - 2 e_0} \over {2 g}}$ \\
 ${\bf e}^{r}_{4} = 2 (0,1)$ & & &
 $x({\bf e}^{r}_{4}) = {{4 - 4 e_0} \over {2 g}}$ \\
\end{tabular}
\label{rota_reci}
\eeq
We see that the most relevant vector is ${\bf e}^{r}_{4}$,
and it is sufficient to keep only the corresponding vertex 
operator in Eq.~(\ref{w_coarse}).

It has been argued in Refs.~\cite{Kondev95,Kondev98} that since the loop
weight does not flow under the renormalisation group, the corresponding
continuum operator $W[{\bf h}({\bf x})]$ must be exactly marginal. In other
words, the screening charge of the Coulomb gas is the electric vector
corresponding to the most relevant vertex operator in the expansion
(\ref{w_coarse}). We therefore set $x({\bf e}^{r}_{4}) = 2$, and as a
consequence the coupling constant $g$ gets fixed as a function of $e_0$
(and hence $n$):
\beq
 g = 1 - e_0 .
\label{coupling}
\eeq  
By virtue of Eq.~(\ref{coupling}), all the critical exponents and the central
charge for the system now gets determined exactly.%
\footnote{The compactification radius of the boson $h_2$ is the repeat lattice
unit vector length ($= 1$) and does not appear explicitly in the dimension
formula (Eq.~(\ref{dimension})) due to our choice of the units of the vectors
${\bf A}$, ${\bf B}$ and ${\bf C}$. But even if such a parameter were
retained, the product of it with $g$ would get fixed as in
Eq.~(\ref{coupling}).}

The central charge of the critical FPL model is obtained by noting that 
there are two bosonic fields $h_1$ and $h_2$, where $h_1$ is free 
and $h_2$ is coupled to the background charge $- 2 e_0$. This is given as 
\cite{Dotsenko84,Kondev95}
\beq
 c = 2+12 x({\bf e}_0,{\bf 0}) = 2 - {6 e_0^2 \over 1 - e_0}.
\label{central}
\eeq

\begin{figure}
\begin{center}
\epsfxsize=11.0cm \epsfysize=11.0cm
\epsfbox{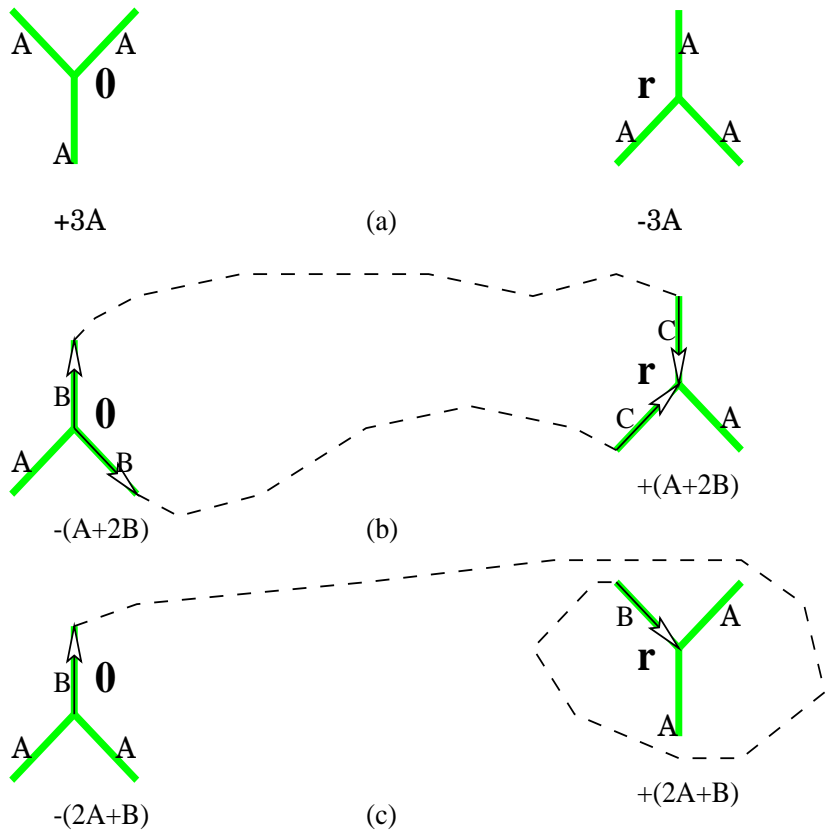}
\caption{Pairs of thermal defects $(a)$, two-string defects $(b)$, and 
one-string defects $(c)$ at the positions ${\bf 0}$ and ${\bf r}$.
The dashed lines denote the paths of open $BCBC\cdots$ strings. In
$(c)$ the defect string winds around the vortex core (see text).}
\label{defects}
\end{center}
\end{figure}

The thermal dimension $x_T$, describing the algebraic decay of the
energy-energy correlator, is linked to a magnetic charge. Namely,
thermal fluctuations breaking the FPL constraint create pairs of
vertices not covered by loops, as shown in Fig.~\ref{defects}$(a)$.
In the height model these are vortices with magnetic
charge ${\bf m} = 3 {\bf A} = (\sqrt{3}, 0)$ and zero electric charge. 
Eqs.~(\ref{dimension}) and (\ref{coupling}) immediately give,
\beq
x_T = {3 \over 2}(1 - e_0).   
\label{thermal}
\eeq 
Note that the electric charges ${\bf e}^R$ and ${\bf e}^r$ that could have
been candidates for the thermal operator are ruled out: the former because
it does not respect the periodicity of the weights, and the latter because
it is irrelevant (and in particular less relevant than (\ref{thermal})).

Similarly, the one-string and the two-string defects, shown in
Fig.~\ref{defects}$(b)$ and $(c)$, correspond to vortices with magnetic
charges ${\bf m}_1 = 2 {\bf A} + {\bf B} = ({\sqrt{3} \over 2}, {1 \over 2}) =
{\bf m}^{R}_2$ and ${\bf m}_2 = {\bf A} + 2 {\bf B} = (0, 1) = - {\bf m}^{R}_1$
respectively.%
\footnote{Note that in order for the height mapping to be globally defined,
the total magnetic charge of the vortices must vanish. In particular,
this implies that the vortex and the anti-vortex that participate in a
two-point correlation function of the thermal or the one-string type
must reside at {\em opposite} sublattices. This is not so for the two-string
type correlation function; see Fig.~\ref{defects}$(b)$.}
At each vortex core there is an additional electric charge
${\bf e}_0$, whose role is to cancel the spurious phase factors arising when
the strings wind around the core \cite{Nienhuis87}. The corresponding
scaling dimensions $x_1$ and $x_2$ thus read
\beq 
 x_1 = x_2 = {(1 - e_0) \over 2} - {{e_0}^{2} \over {2 (1 - e_0)}}.
\label{string12}
\eeq 
The generalisation to $k$-string defects is given in Ref.~\cite{Kondev95}.
Note that the correlation function of a two-string defect propagating from
${\bf 0}$ to ${\bf r}$ is proportional to
the probability of having a loop passing through the points ${\bf 0}$ and
${\bf r}$, and that the two-string dimension is related to the fractal
dimension $D_{\rm f}$ of the loop via the relation \cite{SalDup87,Kondev00},
\beq
 D_{\rm f} = 2 - x_2 . 
\label{fractal}
\eeq

\subsection{FPL model for $w \neq 1$}
\label{FPLw_neq1}

Having reviewed the known results for the FPL model with $w=1$, we now show
how the results get modified for $w \neq 1$, primarily
because of the change of symmetries. We expect that the fixed point at $w=1$
is repulsive, and since the symmetry-breaking effect of taking $w \neq 1$
does not compete with any other parameter (note that we still have $T=0$)
is seems unlikely that the RG flow will halt at any finite value of $w$.
We therefore anticipate a flow towards fixed points at $w=0$ and $w=\infty$
respectively; these expectations are confirmed by our numerical results
and the subsequent analysis.

The fixed point at $w=0$ is trivial, as the only allowed FPL configuration
is the one in which small loops turn around the 1-faces, avoiding all the
$E_0$ edges; cf.~Fig.~\ref{lattice}. Clearly, this state cannot be critical.

The situation for $w\to\infty$ is more interesting. Even though the loops have
to cover all the $E_0$ edges, and stay fully packed, there are many available
configurations. We shall see below that the ideal state graph ${\cal I}$ gets
fragmented, new electric vectors emerge, and that for $|n| \le 2$ the
$w=\infty$ fixed point is in the same universality class as the
critical $q=n^2$ state Potts model.

We begin by examining the ideal state graph in Fig.~\ref{ideal}.
In the limit $w\to\infty$, the two ideal states $(A,B,C)$ and $(A,C,B)$ are
suppressed, as they correspond to colours $A$ lying on top of the $E_0$
edges. We have however checked that for a large lattice with
suitable periodic boundary conditions there exists a sequence of microscopic
updates (permutation of two colours around a closed loop) that link the four
remaining ideal states:
\beq
 (C,B,A) \to (C,A,B) \to (B,A,C) \to (B,C,A) \to (C,B,A) \to \cdots,
\eeq
where all the intermediate configurations in the sequence have the loops
covering all the $E_0$ edges. Indeed, transforming $(C,A,B)$ into $(B,A,C)$
requires interchanging colours $B$ and $C$, and this is easily achieved by
successively flipping the direction of all the $BC$-type loops. Transforming
$(B,A,C)$ into $(B,C,A)$ is more interesting. Flipping first an $AC$-type loop
of length 6 will create a $BC$-type loop of length 18. This latter loop can be
made bigger by flipping further $AC$-type loops that touch its perimeter.
Continuing in this way, the $BC$-type loop will eventually go to the boundary
of the system (and thus annihilate, due to the periodic boundary conditions),
and the region at its interior will transform into the ideal state $(B,C,A)$,
as required.

\begin{figure}
\begin{center}
\epsfxsize=14.0cm \epsfysize=10.0cm
\epsfbox{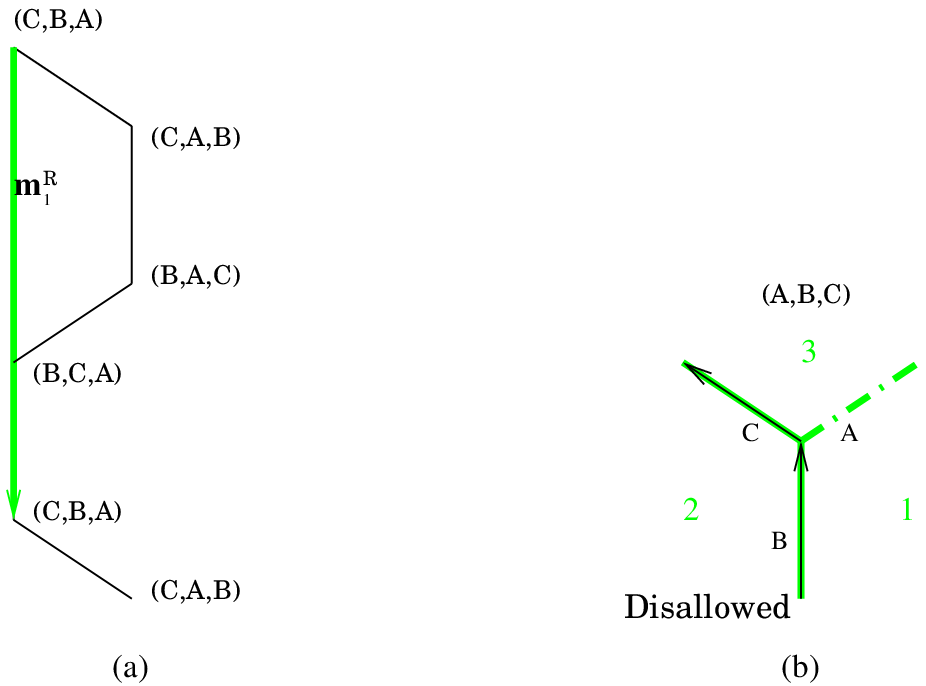}
\caption{$(a)$ The quasi one-dimensional ideal state graph of the
$T=0$, $w\to\infty$ model. $(b)$ The state $(A,B,C)$ is disallowed
as no loop passes over the edge shared by the faces labeled $1$ and $3$.}
\label{ideal2}
\end{center}
\end{figure}

This means on one hand that the two-dimensional ideal state graph in
Fig.~\ref{ideal}$(a)$ becomes fragmented, and on the other hand that
the four remaining possible ideal states forming a valid quasi one-dimensional
ideal state graph. The repeat lattice is now one-dimensional, and is spanned
by the vector ${\bf m}^{R}_1$, as shown in Fig.~\ref{ideal2}$(a)$.

Let us consider what this change implies for the three terms in the
action (\ref{action}). Of the various symmetries used to reduce the number
of parameters in the elastic tensor, only the cyclic exchange of colours
$A \to B \to C$ fails. This means that $K_{11}$ and $K_{22}$ can now
be different. In fact, the structure of ${\cal I}$, as shown in
Fig.~\ref{ideal2}$(a)$, suggests that the fluctuations of the first height
component $h_1$ is suppressed, and $K_{11}\to \infty$. Denoting
$K_{22} = \pi g$ the elastic term therefore takes the form
\beq
 S_{\rm E} = \int {\rm d}^{2}{\bf x} \,
 \exp \left( g \pi ({\bp} h_2)^2 \right).
\label{elasticnew}
\eeq

The argument giving the boundary term $S_{\rm B}$ is unchanged.
So the the bosonic field $h_2$ is again coupled to 
the background charge $- 2 e_0$ given by Eq.~(\ref{e_o}). 

The Liouville term $S_{\rm L}$ is still given by
Eqs.~(\ref{liouville}) and (\ref{w_coarse}), but the microscopic vertex
weights (\ref{w_micro}) are replaced by:
\begin{eqnarray}
 W(C,A,B) &=&   W(B,C,A) = +i {\pi \over 6} e_0 \nn \\
 W(C,B,A) &=&   W(B,A,C) = -i {\pi \over 6} e_0 \nn \\
 W(A,B,C) &=&  +i {\pi \over 6} e_0 - \ln[w] \nn \\
 W(A,C,B) &=&  -i {\pi \over 6} e_0 - \ln[w] 
\label{w_micronew}
\end{eqnarray}
The crucial thing to note is that the invariance under
$120^{\rm o}$ rotations of the ideal states has been lost.

All of the symmetry-related lattices (the ideal state graph, the repeat
lattice, and so on) should now be considered as one-dimensional, as should
the electric and magnetic charges. To see this, consider writing the analogue
of Eq.~(\ref{dimension}) with couplings $K_{11} \neq K_{22}$. In the limit
$K_{11} \to \infty$, the first component of ${\bf e}=(e_1,e_2)$ becomes
immaterial, and any non-zero first component in ${\bf m}=(m_1,m_2)$
would lead to an infinite scaling dimension, whence the corresponding
operator is infinitely irrelevant. The change from a two- to a one-dimensional
lattice will actually modify the computation of electric charges, even if
in some cases they are still reciprocal to the same height periods as
before. To avoid confusion, we shall therefore denote the one-dimensional
electric charges by the letter ${\bf G}$ in the following.

The repeat lattice is now spanned by ${\bf m}^{R}_1 = (0,-1)$. The shortest
vectors in ${\cal R}^*$ and the dimensions of the corresponding vertex
operators read, by Eq.~(\ref{dimension}):
\beq
\begin{tabular}{lll}
${\bf G}^{R}_1 = (0,-1)$ & \qquad \qquad &
$x({\bf G}^{R}_{1}) = {{1 + 2 e_0} \over {2 g}}$ \\
${\bf G}^{R}_2 = (0,1)$ & &
$x({\bf G}^{R}_{2}) = {{1 - 2 e_0} \over {2 g}}$ \\
\end{tabular}
\label{repeat_reci_w}
\eeq
The next-shortest vectors in ${\cal R}^*$ are ${\bf G}^{r}_1 = 2{\bf G}^{R}_1
= 2 (0, -1) $ and ${\bf G}^{r}_2 = 2{\bf G}^{R}_2 = 2 (0,1)$. These are also
the shortest vectors in ${\cal R}_W^*$. To see this, consider the
one-dimensional projection of the ideal state graph shown in
Fig.~\ref{ideal2}. Due to the equality of weights in the first two lines of
Eq.~(\ref{w_micronew}), the weight operator is periodic with half the period
of the repeat lattice.

Interestingly, the electric charges ${\bf G}^r_1$ and ${\bf G}^r_2$ coincide
with ${\bf e}^{r}_1$ and ${\bf e}^{r}_4$ in Eq.~(\ref{rota_reci}); the other
${\bf e}^{r}$'s are now absent. The marginality of the weight operator
therefore still leads to $x({\bf e}^{r}_4) = 2$, and the same fixation of the
coupling, $g = (1 - e_0)$, as in Eq.~(\ref{coupling}).

%

Since there is now only a single bosonic field coupled to the background charge
$- 2 e_0$, we get the central charge as 
\beq
 c = 1 - {6 {e_0}^2 \over 1 - e_0}.
\label{central2}  
\eeq  

We next turn to the thermal exponent. A height defect of magnetic charge
$3 {\bf A}$ is no longer allowed, as this would mean leaving an $E_0$ edge
uncovered. This suggests that the energy operator is linked to an
electric charge instead. We claim further that the appropriate vertex
operator is expected to have the same periodicity as the
vertex weights (\ref{w_micronew}), since the latter encode the energy
of the various microscopic configurations (note that these ``energies''
are purely imaginary in the oriented loop model). The candidate electric
charges are given by Eq.~(\ref{repeat_reci_w}), and they also happen to
be the shortest vectors of ${\cal R}^*$ (and hence correspond to the
most relevant vertex operators). From this argument one should think
that $x_T = x({\bf G}^{R}_{2})$; however, our numerical results
unambiguously show that the correct identification is
\beq
 x_T = x({\bf G}^{R}_{1}) = {{1 + 2 e_0} \over {2 (1 - e_0)}}.
\label{thermal_w}
\eeq
It is seen from Eqs.~(\ref{central2})--(\ref{thermal_w}) that the critical FPL
model at $w \to \infty$ is in the same universality class as the 
$q=n^2$ Potts model. The Coulomb gas of the Potts model is known to
exhibit a so-called charge asymmetry \cite{Nienhuis87,Nienhuis98},
which explains the seemingly paradoxical identification of $x_T$ with
$x({\bf G}^R_1)$, rather than with $x({\bf G}^R_2)$ whose vertex operator
is more relevant.

The two-string defect dimension $x_2$ associated with the magnetic charge
${\bf m}^{R}_1 = (0,-1)$ and the fractal dimension $D_{\rm f}$ are still given
by Eqs.~(\ref{string12}) and (\ref{fractal}).

String defects with any {\em odd} number of strings, however, do not exist at
the fixed point under consideration. This is not surprising, since the
magnetic charges of such defects lead to infinite scaling dimensions, as
explained above. Note also that in the standard formulation of the critical
Potts model as a loop model on the square lattice, strings can be identified
as the boundaries of Fortuin-Kasteleyn clusters \cite{Baxter82},
cf.~section~\ref{sec_staggered}. Thus, in terms of the Potts spins an odd
number of string is meaningless.

Yet another argument for the absence of string defects with an odd number of
strings can be obtained by considering the transfer matrix in the
direction ${\cal T}_\parallel$, in the limit $w\to\infty$. In section
\ref{sec_TM_w_inf} we have showed that the conserved number of occupied
vertical edges in a time-slice must be {\em even}. Clearly, this
rules out the possibility of having states with a single string, and
any number of pairwise connected loop segments.

\begin{figure}
\begin{center}
\epsfxsize=14.0cm \epsfysize=13.0cm
\epsfbox{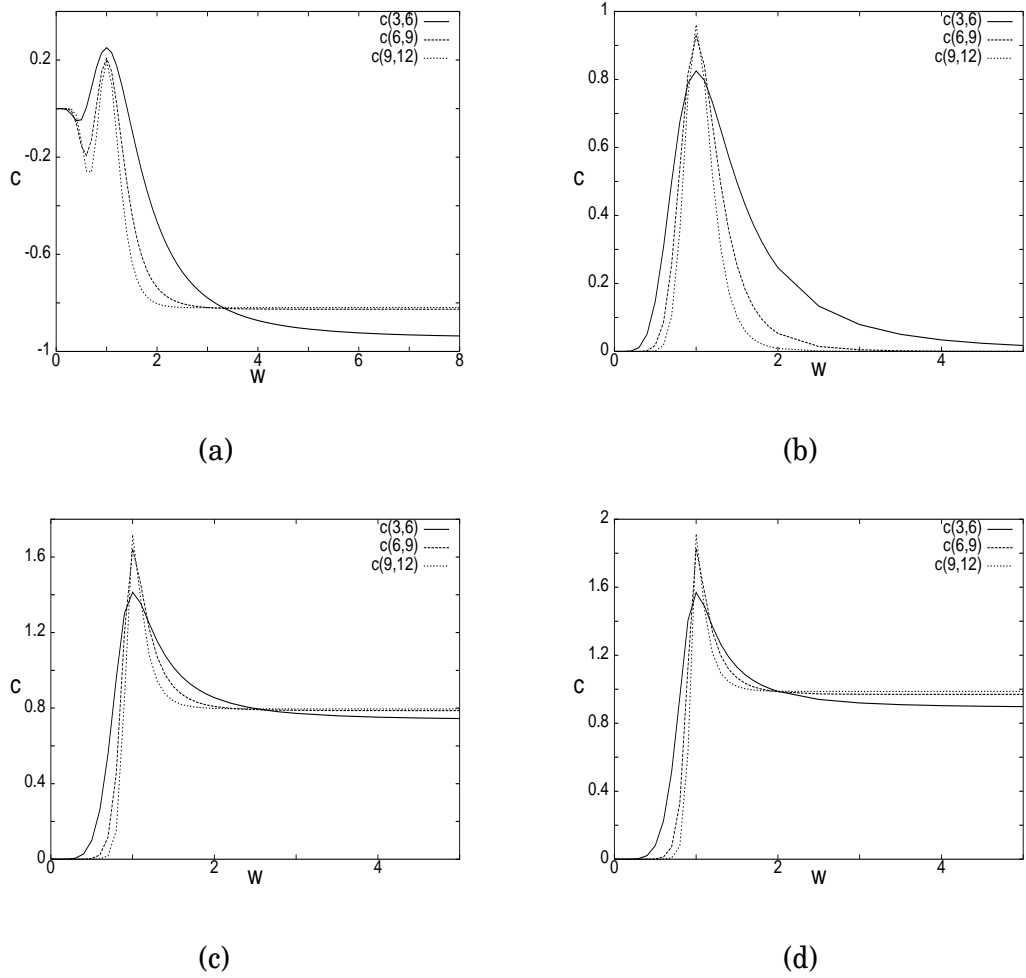}
\caption{The central charge $c$ as a function of $w$ for different $n$: $(a)$ 
$n = 1/2$, $(b)$ $n = 1$, $(c)$ $n = \sqrt{3}$, and $(d)$ $n = 2$.
All the curves are obtained using the two-parameter fits of
Eq.~(\ref{finite_c}).}
\label{cent_w}
\end{center}
\end{figure}

We conclude this section by presenting some numerical evidence for the
correctness of the above predictions. In Fig.~\ref{cent_w} we show the
effective central charge $c_{\rm eff}$ for the $T=0$ model, as a function of
$w$, for various system sizes $L$. The data for different values of $n$ all
agree to show a global maximum at $w=1$. By Zamolodchikov's $c$-theorem
\cite{Zamolo} this means a repulsive fixed point, just as expected. The
repulsive nature is further confirmed by the fact that the peak gets sharper
as the system size increases.

When $w<1$, the RG flow goes all the way down to $w=0$, which is the
location of the trivial fixed point discussed above. Here, $c_{\rm eff}=0$,
indicating non-critical behaviour.

When $w>1$, $c_{\rm eff}$ rapidly attains a plateau,
which extends all the way to $w=\infty$. This indicates an RG flow
towards a fixed point at $w=\infty$. The nature of this fixed point
is attractive, as witnessed by the fact that the plateau gets broader
and flatter with increasing system size.

Finally note that the values of the central charge at the fixed points
$w=1$ and $w=\infty$ are in excellent agreement with the predictions
of Eqs.~(\ref{central}) and (\ref{central2}) respectively.

\begin{table}
\begin{center}
\begin{tabular}{|c|c|c|c|c|c|c|} \hline
$n$ & $L = 6$ & $ L = 9$ & $L = 12$ & $L = 15$ & $L = 18$ & Exact \\ 
\hline
$2.0$	& $0.5592$ & $0.5482$ & $0.5439$ & $0.5417$ & $0.5403$ & $0.5$ 
\\ \hline
$\sqrt{3}$ & $0.7166$ & $0.7265$ & $0.7360$ & $0.7436$ & $0.7496$ & $0.8$ 
\\ \hline
$\sqrt{2}$	& $0.9981$  & $0.9996$ & $0.9999$ & $1.0000$ & $1.0000$ & $1$ 
\\ \hline 
$1.0$	& $1.5286$  & $1.3716$ & $1.3141$ & $1.2898$ & $1.2772$ & $1.25$ 
\\ \hline 
$0.5$	& $0.9466$  & $1.3112$ & $1.3996$ & $1.4588$ & $1.4959$ & $1.5842882$ 
\\ \hline 
\end{tabular} 
\end{center}
\caption{Thermal exponent $x_T$ in the limit $w\to\infty$.}
\label{tab_xtPotts}
\end{table}
 
We have also checked the values of the thermal exponent $x_T$ at the fixed
point $w = \infty$. We here coded our transfer matrix so that only
configurations in which loops covering all the $E_0$ edges were allowed. This
led to a reduction in the number of configurations, and hence exploration of
larger system sizes. We list in Table~\ref{tab_xtPotts}, the exponents
(obtained from Eq.~(\ref{finite_x})) as a function of system size for
different $n$, and compare them with the $x_T$ given by Eq.~(\ref{thermal_w}).
The agreement is very good, in particular when allowing for some
extrapolation of the finite-size data.

\subsection{Finite temperature, $w=1$}
\label{O_n}

Let us now show how the same formalism developed for the FPL models above, can
be used to rederive the known results for the $w=1$ model at finite
temperature \cite{Nienhuis82}. The crucial difference with the FPL model is
that there is now a finite density of the vertex configurations shown in
Fig.~\ref{defects}$(a)$, where three $A$ colours meet. Since these are defects
of magnetic charge ${\Delta {\bf h}} = 3 {\bf A} = (\sqrt{3},0)$, this makes
the two-dimensional height model break down. More precisely, this suggests
that $h_1$ fluctuations are suppressed, and that the model must be described
in terms of the $h_2$ component only. In particular, the ideal state graph
again becomes quasi one-dimensional.

Let us first discuss the generic effective field theory, i.e., at an arbitrary
but small value of $T$. The elastic term and the background term in the action
are given now for the bosonic field $h_2$ by Eqs.~(\ref{elasticnew}) and
(\ref{background}). The local loop weight operator gets modified due to the
new weights of uncovered vertices:
\begin{eqnarray}
 w(C,A,B) &=&   w(B,C,A) = w(A,B,C) = +i {\pi \over 6} e_0 \nonumber \\
 w(C,B,A) &=&   w(B,A,C) = w(A,C,B) = -i {\pi \over 6} e_0 \nn \\
 w(A,A,A) &=&  - \ln[T] 
\label{w_micro_On}   
\end{eqnarray} 
This implies that the coarse-grained loop weight operator (\ref{w_coarse}) is
invariant under all $120^{\rm o}$ rotations. The most relevant electric vector
is again ${\bf G}_2^r = {\bf e}^{r}_4 = 2(0,1)$. The marginality assumption
leads to $x({\bf e}^{r}_4) = 2$, implying $g = 1 - e_0$. This leads to the
central charge being given again by Eq.~(\ref{central2}).

The thermal exponent is linked to an electric charge in the Coulomb gas,
since going to a one-dimensional height has projected to zero the magnetic
charge formerly linked with a temperature-like defect of three $A$ colours.
The most relevant vertex operators have electric charge ${\bf G}^R_1$
and ${\bf G}^R_2$. However, these are ruled out since they do not have
the full symmetry of the weights (\ref{w_micro_On}) with respect to
$120^{\rm o}$ rotations. The next available choices are ${\bf G}^R_1$ and
${\bf G}^R_2$, of which the latter is already in use as the screening operator.
By Eq.~(\ref{rota_reci}) we therefore have
\beq
 x_T = x({\bf G}^{R}_1) = {2(1+e_0) \over 1 - e_0}.
\label{thermalOn}
\eeq 
With $0 \le e_0 < 1$, corresponding to $2 \ge n > -2$, this is indeed
the correct result for dense polymers \cite{Nienhuis82}. Note also
that $x_T > 2$, so that temperature is indeed irrelevant at this fixed point.

Note that the above argument did not in any way use the exactly known
location of the critical point (\ref{Tsolv})${}^-$. The temperature
enters in various non-universal quantities, such as the coefficients
${\tilde W}_{\bf e}$ in Eq.~(\ref{w_coarse}). It therefore may happen
at a {\em particular} value of $T$ that ${\tilde W}_{{\bf G}^r_2} = 0$.
We conjecture that this happens at the critical temperature $T_{\rm c}$,
given by Eq.~(\ref{Tsolv})${}^+$. The screening charge is then
${\bf G}^r_1$, and setting $x({\bf G}^r_1)=2$ we obtain $g=1+e_0$.
Hence
\beq
 c = 1 - \frac{6 e_0^2}{1+e_0},
\label{cdil}
\eeq
and the thermal operator
\beq
 x_T = x({\bf G}^R_2) = \frac{2(1-e_0)}{1+e_0}
\label{xTdil}
\eeq
is now relevant, as it should.
Note that Eqs.~(\ref{cdil})--(\ref{xTdil}) can also be obtained from
Eqs.~(\ref{central2}) and (\ref{thermalOn}) by analytic continuation,
i.e., by taking $-1 < e_0 \le 0$, which once again corresponds to
$-2 < n \le 2$ \cite{Nienhuis82}.

\section{Phase diagram}
\label{Onwings}

\subsection{Fixed points and renormalisation group flows}

\begin{figure}
\begin{center}
\epsfxsize=14.0cm \epsfysize=9.0cm
\epsfbox{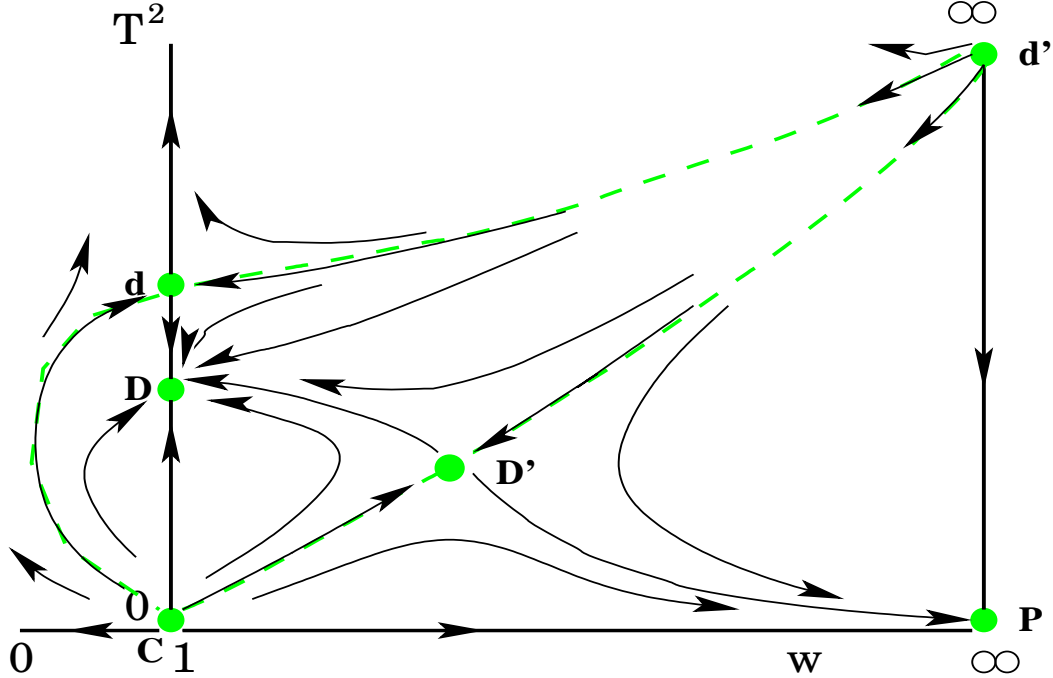}
\caption{The schematic phase diagram for the $O(n)$ model on the hexagonal 
lattice in the $(w,T^2)$ plane at fixed $n \in (0,2)$. Grey filled circles
indicate critical fixed points; for the point at $(\infty,\infty)$ see the
text. The renormalisation group flows are sketched by black arrows.
The interpretation of the various critical points is as follows (see text
for details):
{\sf C} compact loops; {\sf D} dense loops; {\sf d} dilute loops;
{\sf P} Potts model; {\sf D}' tricritical dense loops; {\sf d}' tricritical
dilute loops.}
\label{phasedia}
\end{center}
\end{figure}

In Fig.~\ref{phasedia} we show the schematic phase diagram for the model
(\ref{model}) in the $(w,T^2)$ plane at a fixed value of $n$. 
This diagram is valid for $0 \leq n < 2$. The case $n=2$ is slightly
special and will be discussed separately in section~\ref{sec_special}.

The critical fixed points {\sf C} (compact loops), {\sf D} (dense loops) and
{\sf d} (dilute loops), were previously known and have been discussed in the
introduction. The direction of the RG flows for $w=1$ follow from the
relevance (resp.~irrelevance) of $x_T$ at {\sf C} and {\sf d} (resp.~at {\sf
D}). Note also the agreement with the $c$-theorem \cite{Zamolo}, according to
which the effective central charge decreases along RG flows.

In section~\ref{FPLw_neq1} we have demonstrated that for $T=0$ there is
a flow from {\sf C} to another fixed point {\sf P} (Potts model) situated
at $w=\infty$.

It should be noted that the points {\sf D} and {\sf P} are both generic in the
sense that their Coulomb gas is one-dimensional with no tunable parameters.
(In particular, their central charges coincide.) This would seem to indicate
that these points should be attractive in all directions. In the case of
{\sf D} this is supported by the irrelevance of the temperature perturbation,
i.e., $x_T > 2$ in Eq.~(\ref{thermalOn}). In the case of {\sf P}, we have seen
in section~\ref{FPLw_neq1} that $1/w$ is irrelevant. It is also true that
$T$ is irrelevant. This can be seen, on one hand, microscopically by noticing
that the $w\to\infty$ limit enforces full packing for any finite $T$. On
the other hand, one has formally $x({\bf 0},3 {\bf A}) = \infty$ in
Eq.~(\ref{dimension}), cf.~the remark after Eq.~(\ref{w_micronew}).
Thus finite $T$ is indeed infinitely irrelevant at {\sf P}.%
\footnote{This statement is seemingly contradicted by the fact that
$x_T < 2$ at {\sf P} [see Eq.~(\ref{thermal_w})]. However, we remind that
the notation $x_T$ refers to the scaling dimension of the energy
operator, which is related to the next-largest translationally invariant
eigenvalue of the transfer matrix. This does not necessarily coincide with
the operator that couples to the temperature variable $T$. Thus,
the exponent $x_T$ of Eq.~(\ref{thermal_w}) must correspond to one
of the nine independent way of leaving the manifold (\ref{hex_tri_map})
which embeds the model (\ref{model}) within the more general
triangular-lattice O($n$) model.}

Accepting the attractive nature of {\sf D} and {\sf P} in all directions
in the $(w,T^2)$ plane as a working hypothesis,
there must be a curve emanating from {\sf C} that separates their respective
basins of attraction. For the qualitative reasons given towards the end
of section~\ref{sec_staggered}, it seems reasonable to suppose there will
be an RG flow along this curve, going from {\sf C} towards a new
multicritical point {\sf D'} situated at finite $T$ and $w$.

The simplest way of obtaining a phase diagram that is consistent
with the properties listed above is to suppose the existence of yet another
fixed point {\sf d'} at infinite $T$ and $w$, which is repulsive in all
directions in the $(w,T^2)$ plane. This yields then the conjectured phase
diagram shown in Fig.~\ref{phasedia}. The limit $w,T\to\infty$ should be
taken with $\lambda \equiv T^2/w$ fixed, since uncovering an $E_0$ edge
will necessarily lead to the formation of two empty vertices. Taking
$\lambda \to 0$ we recover an FPL model, i.e., the point {\sf P}. On the
other hand, $\lambda \to \infty$ yields a completely empty configuration.
The hypothesis is thus that there exists a critical fixed point {\sf d'}
situated at some finite value $\lambda = \lambda_{\rm c}$.

\begin{figure}
\begin{center}
\epsfxsize=11.0cm \epsfysize=8.0cm
\epsfbox{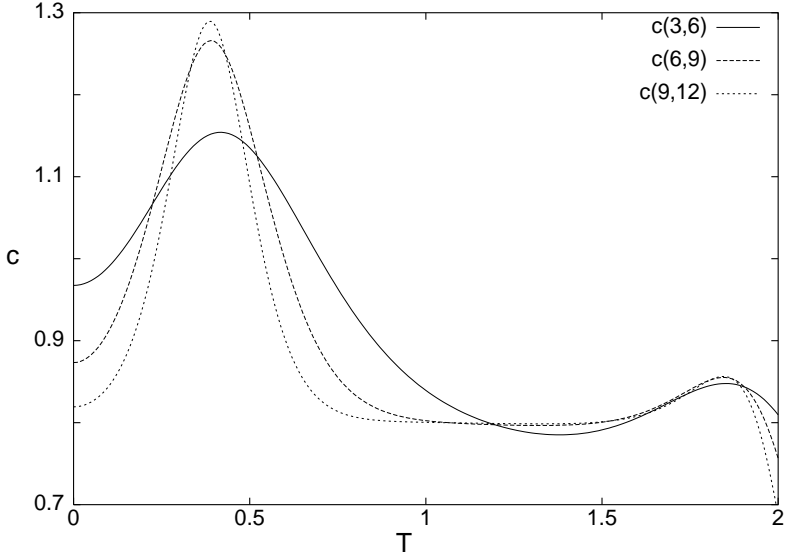}
\caption{Plot of $c_{\rm eff}$ versus $T$ at $n = \sqrt{3}$ and $w = 1.60$,
for various system sizes $L$. The two maxima are at $T^-(w) \simeq 0.39$ and
$T^+(w) \simeq 1.84$.}
\label{c_T}
\end{center}
\end{figure}

\subsection{Universality classes and numerical results}

To verify the hypotheses made above---and to establish the universality
classes of the proposed fixed points {\sf D'} and {\sf d'}---we proceed
to make some numerical checks.

To get some guidance, we show in Fig.~\ref{c_T} a plot of the effective
central charge $c_{\rm eff}$ (i.e., the value of $c$ obtained by fitting
the numerically obtained free energies to Eq.~(\ref{finite_c}) for two
different system sizes $L$) as a function of $T$, at fixed $n = \sqrt{3}$ and
$w = 1.60$. At $T=0$, with increasing system size $L$ we have $c_{\rm eff} \to
4/5$, in agreement with the prediction for the point {\sf P},
cf.~Eq.~(\ref{central2}). At $T=T^-(w) \simeq 0.39$, one observes a maximum in
$c_{\rm eff}$. This is the signature of the line of RG flows that separates
the basins of attraction of {\sf D} and {\sf P} in Fig.~\ref{phasedia}. This
is followed by a plateau in $c_{\rm eff}$ signaling the basin of attraction of
{\sf D}. The broadening of this plateau with increasing $L$ is consistent with
the attractive nature of {\sf D}. Moreover, the plateau value $c_{\rm eff}
\simeq 4/5$ is in agreement with the prediction for {\sf D}. At $T=T^+(w)
\simeq 1.84$, the plateau terminates in a second maximum. This is the
signature of the line of RG flows that separates the basins of attraction of
{\sf D} and the trivial fixed point at $(w,T^2) = (0,\infty)$. For $T > T^+$,
$c_{\rm eff}$ levels off to zero, signaling non-critical behaviour. Note that
the sharpening of the two maxima at $T^\pm(w)$ with increasing $L$ is
consistent with the corresponding RG flows' being repulsive in the temperature
direction.

\begin{table}
\begin{center}
\begin{tabular}{|c|c|c|c|l|} \hline
$w$ & $c(3,6)$ & $c(6,9)$ & $c(9,12)$ & ${T}^-(w)$ \\ \hline
$1.60$	& $1.1541$ & $1.2660$ & $1.2895$ & $0.39 \pm 0.01$ \\ \hline
$2.00$	& $1.1847$ & $1.2762$ & $1.2921$ & $0.56 \pm 0.01$ \\ \hline
$3.30$	& $1.2206$ & $1.2838$ & $1.2943$ & $0.99 \pm 0.01$ \\ \hline
$3.42$	& $1.2204$ & $1.2839$ & $1.2943$ & $1.028 \pm 0.001$ \\ \hline
$3.50$	& $1.2200$ & $1.2838$ & $1.2942$ & $1.05 \pm 0.01$ \\ \hline
$4.00$	& $1.2158$ & $1.2832$ & $1.2941$ & $1.19 \pm 0.01$ \\ \hline
$6.00$	& $1.1919$ & $1.2786$ & $1.2930$ & $1.66 \pm 0.01$ \\ \hline
\end{tabular} 
\end{center}
\caption{Low-temperature maxima of $c_{\rm eff}(T)$, as a function of system
size $L$,
for various values of $w$ and fixed $n=\sqrt{3}$. We also show the estimates
of $T^-(w)$, extrapolated to the $L\to\infty$ limit.}
\label{tab_T-n3}
\end{table}

The curves $c_{\rm eff}(T)$ have a similar functional form (two maxima
separated by a plateau) for other values of $w$. This enables us to
determine the curves $T^\pm(w)$ for any fixed value of $n$. In
Table~\ref{tab_T-n3} we show the results for $T^-(w)$ at $n=\sqrt{3}$.
Extrapolations to the limits $w\to 1$ and $w\to\infty$ confirms that
this RG separatrix connects the points {\sf C} and {\sf d'} in
Fig.~\ref{phasedia}. Actually, we expect $T^-(w)$ to diverge as
$w^{1/2}$ when $w\to\infty$, since {\sf d'} is located at finite
$\lambda = T^2/w$. Examining now the variation of
$c_{\rm eff}\left(T^-(w)\right)$, we observe a maximum at
\beq
 w_{\rm c} = 3.42 \pm 0.01 \mbox{\ \ and \ \ }
 T^-(w_{\rm c}) = 1.028 \pm 0.001 \qquad (n=\sqrt{3}).
\eeq
This is our estimate of the location of the point {\sf D'}, here for
$n=\sqrt{3}$. Note that the maximum gets broader with increasing $L$,
in agreement with the attractive nature of {\sf D'} along the separatrix;
see Fig.~\ref{phasedia}.
Extrapolating the corresponding central charge to $L\to\infty$ we obtain
\beq
 c_{\sf D'}(n=\sqrt{3}) = 1.304 \pm 0.005.
\label{cD'n3}
\eeq


We have repeated this analysis for $n=\sqrt{2}$, finding similar qualitative
results. The point {\sf D'} is here located at
\beq
 w_{\rm c} = 3.60 \pm 0.01 \mbox{\ \ and \ \ }
 T^-(w_{\rm c}) = 1.030 \pm 0.001 \qquad (n=\sqrt{2}),
\eeq
and the central charge reads
\beq
 c_{\sf D'}(n=\sqrt{2}) = 1.004 \pm 0.005.
\label{cD'n2}
\eeq

The mapping from the model (\ref{model}) to the triangular-lattice
O($n$) model \cite{Nienhuis98} exhibited in section~\ref{sec_decimation}
makes it plausible that the points {\sf D'} and {\sf d'} should coincide
with critical behaviour found in the latter model. We therefore conjecture
that the universality class of {\sf D'} is that of a superposition of
dense loops and a critical Ising model, referred to as branch 5
in Ref.~\cite{Nienhuis98}. This means
\bea
   c &=& \frac32 - \frac{6 e_0^2}{1-e_0}. \label{D'c}\\
   x_T &=& 1, \label{D'xT}
\eea
since the thermal exponent of the Ising model is always more relevant than
that of dense loops.

This conjecture for $c$ is in excellent agreement with Eqs.~(\ref{cD'n3})
and (\ref{cD'n2}). The numerical estimates for $x_T(L)$ read, for $n=\sqrt{3}$,
\beq
 x_T(6) = 0.8622, \quad
 x_T(9) = 0.8782, \quad
 x_T(12) = 0.8880.
\eeq
For $n=\sqrt{2}$ we find
\beq
 x_T(6) = 0.9920, \quad
 x_T(9) = 0.9973, \quad
 x_T(12) = 0.9986.
\eeq
The agreement with Eq.~(\ref{D'xT}) is reasonable for $n=\sqrt{3}$ (the
finite-size corrections are strong), and excellent for $n=\sqrt{2}$.

\begin{table}
\begin{center}
\begin{tabular}{|c|c|c|c|l|} \hline
$w$ & $c(3,6)$ & $c(6,9)$ & $c(9,12)$ & ${T}^+(w)$ \\ \hline
$1.04$	& $0.859630$ & $0.856344$ & $0.856625$ & $1.6079 \pm 0.0001$ \\ \hline
$1.08$	& $0.859414$ & $0.856323$ & $0.856622$ & $1.6277 \pm 0.0001$ \\ \hline
$1.12$	& $0.859065$ & $0.856291$ & $0.856616$ & $1.6471 \pm 0.0001$ \\ \hline
$1.20$	& $0.858011$ & $0.856190$ & $0.856599$ & $1.6847 \pm 0.0001$ \\ \hline
$1.60$	& $0.847764$ & $0.855149$ & $0.856392$ & $1.845 \pm 0.001$ \\ \hline
$2.00$	& $0.833671$ & $0.853567$ & $0.856042$ & $1.967 \pm 0.001$ \\ \hline
$4.00$	& $ -- $     & $0.843420$ & $0.853420$ & $2.456 \pm 0.005$ \\ \hline
$6.00$	& $ -- $     & $0.836064$ & $0.850333$ & $2.80 \pm 0.02$ \\ \hline
\end{tabular} 
\end{center}
\caption{High-temperature maxima of $c_{\rm eff}(T)$, as a function of
system size $L$,
for various values of $w$ and fixed $n=\sqrt{3}$. We also show the estimates
of $T^+(w)$, extrapolated to the $L\to\infty$ limit.}
\label{tab_T+n3}
\end{table}

Following the second maximum in Fig.~\ref{c_T} allows us similarly to
trace out the curve $T^+(w)$. The result is shown in Table~\ref{tab_T+n3}.
This separatrix is seen to connect the points {\sf d} and, quite
plausibly, {\sf d'}. Once again we expect $T^+(w)$ to behave as $w^{1/2}$
when $w\to\infty$, and this is confirmed by the data in Table~\ref{tab_T+n3}.
Furthermore, the variation of $c_{\rm eff}$ with $L$
is now consistent with {\sf d} being attractive along the separatrix,
exactly as shown in Fig.~\ref{phasedia}.

We have also followed the maxima in Fig.~\ref{c_T} into the region
$0 \le w < 1$. We find that for a certain value $w=w_0$ the two maxima
coalesce, and for $w<w_0$ the model is non-critical for any $T$. 
The interpretation in terms of RG flows is shown in Fig.~\ref{phasedia}.
Note in particular that it is possible to flow directly from {\sf C}
to {\sf d}, by perturbing {\sf C} by a fine-tuned linear combination
of the operators $w$ and $T^2$, and that the dense-loop phase (i.e.,
the basin of attraction of {\sf D}) only exists for $w>w_0$.
For $n=\sqrt{2}$, we have determined $w_0=0.32 \pm 0.01$, corresponding
to a temperature $T_0=0.92 \pm 0.01$.

Let us finally examine the properties of the point {\sf d'}. As discussed
above, we must take the limit $w,T\to\infty$ with $\lambda=T^2/w$ fixed
in order to see competition between FPL configurations (with weight
$\sim w^{N/2}$, where $N$ is the number of vertices of the lattice)
and empty space (with weight $T^N$). In this limit, the participating
configurations consist of loops along which $E_0$ and non-$E_0$ edges
alternate. An $E_0$-edge not covered by a loop carries a weight $\lambda$.
This reformulation allowed us to write a transfer matrix for this limit,
with parameters $\lambda$ and $n$ (the loop weight).

\begin{figure}
\begin{center}
\epsfxsize=11.0cm \epsfysize=8.0cm
\epsfbox{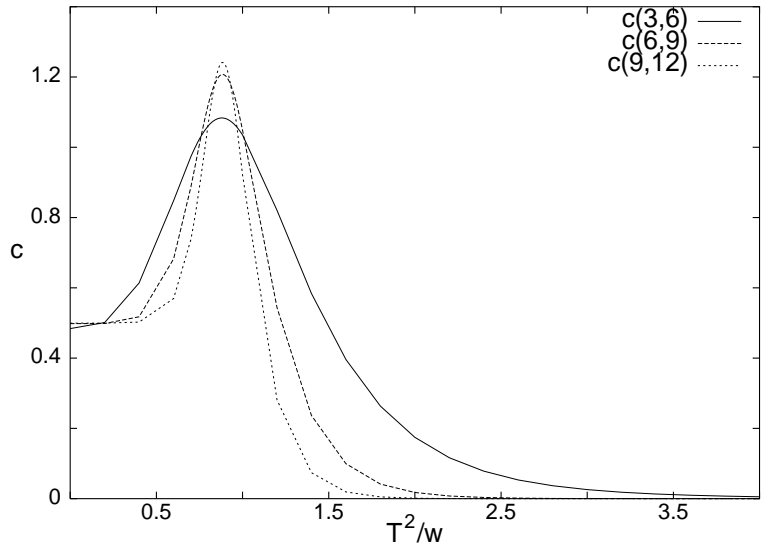}
\caption{$c_{\rm eff}$ as a function of $\lambda=T^2/w$ with
$T,w\to\infty$ and fixed $n = \sqrt{2}$.} 
\label{c_lambda}
\end{center}
\end{figure}

In Fig.~\ref{c_lambda} we display a plot of $c_{\rm eff}(\lambda)$ at
$n = \sqrt{2}$. As $\lambda \to 0$ we recover the point {\sf P} with $c=1/2$,
and in the large-$\lambda$ limit non-critical behaviour is signaled by
$c_{\rm eff}=0$. A maximum at $\lambda=\lambda_{\rm c}\simeq 0.88$
corresponds to the critical behaviour of the point {\sf d'}. The sharpening
of the maximum with increasing $L$ shows its repulsiveness in the variable
$\lambda-\lambda_{\rm c}$.

\begin{table}
\begin{center}
\begin{tabular}{|c|c|c|c|c|c|} \hline
$n$        & $c(3,6)$& $c(6,9)$&$c(9,12)$&$c(12,15)$& $\lambda_{\rm c}$ \\ \hline
$2$        & $1.304$ & $1.409$ & $1.411$ & $ -- $  & $0.79 \pm 0.01$ \\ \hline
$\sqrt{3}$ & $1.222$ & $1.344$ & $1.367$ & $1.369$ & $0.83 \pm 0.01$ \\ \hline
$\sqrt{2}$ & $1.084$ & $1.207$ & $1.241$ & $1.253$ & $0.88 \pm 0.01$ \\ \hline
$1$	   & $0.825$ & $0.928$ & $0.961$ & $0.976$ & $1.00 \pm 0.01$ \\ \hline
\end{tabular} 
\end{center}
\caption{Maxima of $c_{\rm eff}(\lambda)$ in the limit
$w,T\to\infty$ with fixed $\lambda=T^2/w$, as a function of system size $L$,
for various values of $n$. We also show the estimates of the
critical point $\lambda_{\rm c}$, extrapolated to the $L\to\infty$ limit.}
\label{tab_Tinf}
\end{table}

In Table~\ref{tab_Tinf} we locate the point {\sf d'} for several values of
$n$ and we give the corresponding values of the central charge. 
The latter prompt us to conjecture
that the universality class of {\sf d'} is that of a superposition of
dilute loops and a critical Ising model, referred to as branch 4
in Ref.~\cite{Nienhuis98}. In other words,
\beq
   c = \frac32 - \frac{6 e_0^2}{1+e_0}. \label{d'c}
\eeq
We have verified that the point {\sf d'} can be detected as in
Fig.~\ref{c_lambda} for $0 \le n < 2$. 

\begin{table}
\begin{center}
\begin{tabular}{|c|c|c|c|c|} \hline
$n$        & $x_T(3)$& $x_T(6)$&$x_T(9)$ &$x_T(12)$ \\ \hline
$2$        & $0.7106$& $0.4975$& $0.4200$& $0.3685$ \\ \hline
$\sqrt{3}$ & $0.8509$& $0.6094$& $0.5411$& $0.4998$ \\ \hline
$\sqrt{2}$ & $1.1245$& $0.7805$& $0.7203$& $0.6886$ \\ \hline
$1$	       & $ -- $  & $1.1082$& $1.0435$& $1.0237$ \\ \hline
\end{tabular} 
\end{center}
\caption{Thermal exponent $x_T$ at the point {\sf d'},
as a function of system size $L$,
for various values of $n$.}
\label{tab_xTinf}
\end{table}

As to the thermal exponent, one would expect $x_T$ to be the least of
the values given by Eqs.~(\ref{xTdil}) and (\ref{D'xT}).
Curiously, this is not brought out by the data of Table~\ref{tab_xTinf}.
And the numerical data reported in Table 3 of Ref.~\cite{Nienhuis98}
for branch 4 of the triangular-lattice O($n$) model agree neither with
the theoretical expectation, nor with the data of Table~\ref{tab_xTinf}.

\subsection{Coulomb gas arguments for multicritical behaviour}
\label{sec_CG_multi}

Among the various critical branches identified for the O($n$) models
on the square \cite{Nienhuis92} and the triangular \cite{Nienhuis98}
lattices, and for the hexagonal-lattice O($n$) model in a staggered
field [here], four appear to be generic. Those are dense
({\sf D}) and dilute ({\sf d}) loops, and their respective superpositions
with a critical Ising model ({\sf D'} resp.~{\sf d'}).

In Ref.~\cite{Kondev99}, it was explained how these four branches may
be recovered in a more general setting from Coulomb gas arguments.
The reasoning given in Ref.~\cite{Kondev99} applies to the square lattice,
but can be readily adapted to the triangular-lattice O($n$) model
defined in Fig.~\ref{trivertex}.

Indeed, consider a generalisation of the triangular-lattice O($n$) model
in which loops can be either black or white, with respective weights
$n_{\rm b}$ and $n_{\rm w}$. Each edge is covered by either a black or
a white loop. At a vertex, two loops can cross only if they have different
colours. This can again be illustrated by Fig.~\ref{trivertex}, but the
vertices marked $\rho_2$ and $\rho_4$ now each correspond to two
distinct vertices depending on how the white loops are interconnected.
Clearly, the original model is recovered by setting $n_{\rm w}=1$ and
renormalising the weights of the ``doubled'' vertices.

The Coulomb gas is once again based on an interface model on the dual lattice.
After orienting each loop, two-dimensional microscopic heights are defined,
depending on the orientation and colour of each edge. On the square lattice,
this construction was based on an edge-colouring model \cite{Kondev99} (like
in section~\ref{sec_CG}) which defined loops as alternating sequences of two
colours. This is not possible on the triangular lattice, as the lattice is not
bipartite and loop lengths are not necessarily even \cite{Jacobsen99}. The
resulting height model, however, still exists and has the crucial property
that the two loop colours define independent directions in the height space%
\footnote{This is in contrast with the appearance of an ``unexpected'' extra
height component in the FPL models on the hexagonal \cite{Kondev95} and square
\cite{Kondev98} lattices, which are both linked to edge-colouring models.}
(this construction---as well as the detailed arguments---is exactly the same
as in Ref.~\cite{Kondev99}).

The arguments given in Ref.~\cite{Kondev99} thus go through, and show that
the two loop colours lead to two independent one-component Coulomb gases
in the continuum limit. Therefore, four critical branches arise depending
on whether each loop colour is in the dense or in the dilute phase.
In particular, setting $n_{\rm w}=1$, the white loops add either
$0$ or $1/2$ to the central charge, depending on whether they are in the
dense or the dilute phase. Indeed, the dilute O($n_{\rm w}=1$) model is
nothing but a critical Ising model.

\subsection{The case of $n = 2$}
\label{sec_special}

The loop weight $n = 2$ is somewhat special. At $w = 1$, the
critical points {\sf D} and {\sf d} become identical and describe the
Kosterlitz-Thouless point of the XY model. In the usual O($2$) model,
this point is the critical end-point of a line of low-temperature fixed
points along which the exponents vary continuously.

Interestingly, this feature is not present in the loop model \cite{Blote89};
we recall that the latter is defined by a truncation of the full O($n$) model,
which is not innocuous at low temperature. Thus, in the whole region $0 < T <
T^\pm(w=1)$, the critical exponents are constant and equal to their values at
$T^\pm(w=1)$.%
\footnote{Notice that in this respect the phase diagram reproduced in
Ref.~\cite{Wu00} is slightly misleading.}
In particular, the thermal exponent is marginally irrelevant (resp.~marginally
relevant) for temperatures slightly below (resp.~above) $T^\pm(w=1)$.

For $w > 1$, the lines $T^{-}(w)$ and $T^{+}(w)$ are present,
just as for other $n < 2$. In particular we find the {\sf D'}
critical point at $w_{\rm c} = 2.9 \pm 0.1$ and $T^{-}(w_{\rm c}) = 0.91 \pm
0.04$. Its central charge agrees well with the prediction of Eq.~(\ref{D'c}),
which reads $c=3/2$.

Moreover, from Eq.~(\ref{d'c}) we also expect a value of $c = 3/2$ at the
point {\sf d'}; the numerical data in Table~\ref{tab_Tinf} are consistent with
this expectation. This opens for the interesting possibility that the part of
the curve $T^-(w)$ that connects {\sf D'} and {\sf d'} may be a {\em line of
fixed points} with $c=3/2$.

The numerical data for $c$, $x_T$ and $x_1$ shown in Table~\ref{tab_T-n2} are
not inconsistent with this possibility. While $c$ and $x_1$ are fairly
constant along the proposed line of fixed points, $x_T$ steadily decreases
beyond $w > w_{\rm c}$ (also note that the $w\to\infty$ limit of Table
\ref{tab_T-n2} is consistent with the value of $x_T$ given in
Table~\ref{tab_xTinf}). However, if this scenario were indeed true,
one should be able to identify an exactly marginal operator at {\sf D'}.
We plan to investigate this issue further in the future.

\begin{table}
\begin{center}
\begin{tabular}{|c|c|c|c|c|c|c|c|l|} \hline
$w$ & $c(6,9)$  & $c(9,12)$ & $x_1(9)$ & $x_1(12)$ & $x_T(9)$ & $x_T(12)$ & $T^{-}(w)$ \\ \hline
$1.6$	& $1.46271$ & $1.48616$ & $0.2277$ & $0.2249$ & $0.7151$ & $0.7231$ & $0.43 \pm 0.01$ \\ \hline
$2.0$	& $1.47149$ & $1.48938$ & $0.2303$ & $0.2282$ & $0.7226$ & $0.7283$ & $0.60 \pm 0.01$ \\ \hline
$2.9$	& $1.47670$ & $1.49182$ & $0.2401$ & $0.2411$ & $0.7136$ & $0.7150$ & $0.91 \pm 0.01$ \\ \hline
$4.0$	& $1.47452$ & $1.49078$ & $0.2401$ & $0.2407$ & $0.7051$ & $0.7067$ & $1.23 \pm 0.01$ \\ \hline
$6.0$	& $1.46613$ & $1.48739$ & $0.2434$ & $0.2441$ & $0.6782$ & $0.6786$ & $1.70 \pm 0.01$ \\ \hline
$10.0$	& $1.45331$ & $1.48173$ & $0.2645$ & $0.2703$ & $0.6186$ & $0.6107$ & $2.40 \pm 0.01$ \\ \hline
$40.0$	& $1.43237$ & $1.46268$ & $0.3106$ & $0.3169$ & $0.4967$ & $0.4655$ & $5.46 \pm 0.01$ \\ \hline
\end{tabular} 
\end{center}
\caption{Estimates of $c_{\rm eff}$, $x_1$ and $x_T$ along the
$T^{-}(w)$ curve at $n=2$, for various system sizes.}
\label{tab_T-n2}
\end{table}

\subsection{Hard hexagons and hard triangles: $n>2$}

When $w=1$, taking the limit $T,n\to\infty$ with fixed $z=n/T^6$ transforms
the model (\ref{model}) into Baxter's model of hard hexagons
\cite{hardhex,Baxter82}. The role of the hexagons is played by the loops of
length six, which may surround any of the hexagonal-lattice faces. As a
function of their fugacity $z$ there is a solid-to-liquid transition at
$z=z_{\rm c}=(11+5\sqrt{5})/2$ at which the model is exactly solvable.
For $z\to\infty$ there is a coexistence of three completely ordered
phases, in which the loops are fully packed and all surround faces with
the same label. This may lead one to expect some connection with a three-state
Potts model, along the lines of section~\ref{sec_staggered}. And actually
the exact solution at $z=z_{\rm c}$ shows \cite{hardhex} that the
critical exponents of the hard hexagon model coincide with those of the
three-state Potts model.

More recently, it was shown numerically \cite{Wu00} that this critical
behaviour extends to a critical curve $z_{\rm c}(n)$ existing for any $n>2$.
In particular, as $n \to 2_+$ one has $z_{\rm c}(n) \to \infty$ (or
$T\to 0$). This remarkable finding means that the $Z_3$ symmetry is felt
even by the ``soft'' particles (loops of length longer than six) at
finite $n$. The results of Ref.~\cite{Wu00} can be interpreted in
RG terms by assuming that there is a curve of RG flows along the curve
$z_{\rm c}(n)$ directed towards the critical fixed point at $n=\infty$.
Note that an RG flow changing the symmetry related parameter $n$ is
possible exactly because at $n>2$ the loop weight operator is no longer
marginal.

In section~\ref{sec_staggered} we have discussed how taking $w>1$ breaks
three-phase coexistence down to two-phase coexistence. One is thus led
to conjecture that for $n>2$ there will exist a critical manifold in
the $w>1$ half space that belongs to the Ising universality class
($Z_2$ type symmetry).

Let us consider more closely taking first $w\to\infty$, and then
$T,n \to\infty$ with fixed $z=n/T^6$. Decimating the 2-faces as in
section~\ref{sec_decimation} we recover a model of {\em hard triangles}.
These triangles are loops of length three that surround the faces of the
triangular lattice; they may touch at the vertices, but not share an edge.
The fugacity of a triangle is $z$. The two completely ordered states
correspond to having all the triangles surround the faces labeled
1 or 3, respectively. It seems natural to conjecture that coexistence
of these two states may lead to critical behaviour of the Ising type.%
\footnote{In order to compare with the conventional formulation of models
of hard particles, it is useful to consider the dual lattice.
The particles of fugacity $z$ now occupy the vertices of the hexagonal
lattice, and the hard-core constraint is that no two neighbouring vertices
can be occupied. In this sense, the model is one of hard {\em squares}
whose centres live on the hexagonal lattice. Note that this is different
from the hard squares model where the particles live on the square lattice.
Neither of these models appear to have been exactly solved.}

We have checked these conjectures numerically. For any $n>2$ we find that
at $w=1$ there exists a critical temperature $T_{\rm c}(w=1)$ with a
$c=4/5$ behaviour, consistent with the findings of Ref.~\cite{Wu00}.
For $w>1$, this extends into a critical curve $T_{\rm c}(w)$, with an
RG flow directed towards larger $w$ that eventually ends up in a $c=1/2$
critical fixed point in the $w\to\infty$ limit. The location of this fixed
point is such that $\lambda = T^2/w$ tends to a constant. This
flow from $c=4/5$ to $c=1/2$ divides the phase diagram into a low-temperature
ordered (solid) state and a high-temperature disordered (liquid) state;
there appears to be no further critical points.

\section{Loops on the triangular lattice}
\label{sec_tri}

We finally consider several non-standard FPL models on the triangular lattice.
These loop models exhibit different critical behaviour than the
standard FPL model on the triangular lattice, which is in the
universality class of dense loops \cite{Batchelor96,Jacobsen99}.

\begin{figure}
\begin{center}
\epsfxsize=6.0cm \epsfysize=4.0cm
\epsfbox{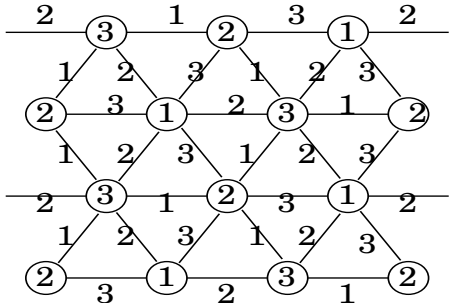}
\caption{Triangular lattice obtained by superposing three shifted
hexagonal lattices. The vertices (here surrounded by circles for clarity)
and edges are labeled by integers $k=1,2,3$ as shown.}
\label{fig_3hex}
\end{center}
\end{figure}

Consider constructing a triangular lattice by superposing three shifted
hexagonal lattices, as shown in Fig.~\ref{fig_3hex}.
The vertices of the triangular lattice are labeled $k=1,2,3$ so that
no two adjacent vertices have the same label. Let $(k_1,k_2,k_3)$ be any
permutation of $(1,2,3)$. Then the $k_1$'th hexagonal lattice consist
of those edges of the triangular lattice that join vertices labeled
$k_2$ and $k_3$. Conversely, at a triangular-lattice vertex labeled
$k_1$, the hexagonal lattices labeled $k_2$ and $k_3$ share a common vertex.
Each edge of the triangular lattice belongs to exactly one of the
three hexagonal lattices.

\subsection{The RGB model}
\label{sec_RGB_model}

\begin{figure}
\begin{center}
\epsfxsize=14.0cm \epsfysize=6.0cm
\epsfbox{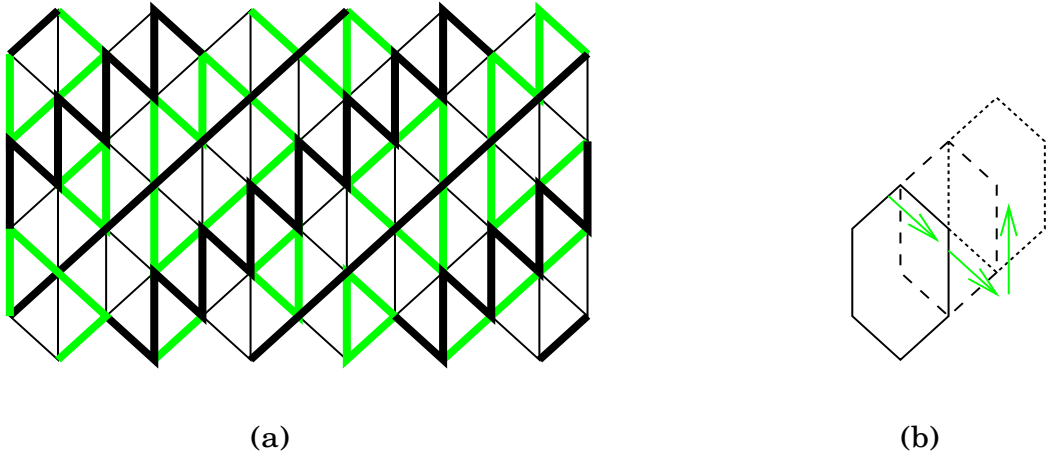}
\caption{$(a)$ The thick grey loops are the RGB loops. $(b)$ The three 
hexagonal lattices are shown by the continuous, broken, and dotted lines
respectively. Following successive dimers along edges on these three lattices,
as shown by the arrows, one gets an RGB loop.}
\label{RGB}
\end{center}
\end{figure}

The first model we consider is known as the red-green-blue (RGB) model, and
was introduced and studied in Refs.~\cite{BenjSch,Wilson02}. It is obtained by
superposing independent dimer coverings of each of the three hexagonal
lattices. We shall here study the more general case where the hexagonal
lattices are endowed with three-colouring configurations,
cf.~section~\ref{sec_3C_model}, each $BC$-type loop being weighted by $n$.
Dimers correspond to colours $A$, and to weigh each dimer configuration
equally we simply set $n=1$. The dimers now form closed loops---the RGB
loops---which are each weighted by unity. Apart from being fully packed, the
RGB loops have the interesting properties that their length is always a
multiple of three, and that they never make $\pi/3$ turns. An example is shown
by the thick grey loops in Fig.~\ref{RGB}$(a)$ (disregard for the
moment the other two kinds of loops shown).

In Ref.~\cite{Wilson02} a one-dimensional height model was associated
with the RGB loops as follows. Imagine starting at a definite vertex of
the triangular lattice, and traversing the RGB loop that goes through
this vertex by moving in the clockwise direction. Each edge on the trajectory
is a dimer on one of the three hexagonal lattices, and when tracing out
the loop we note sequentially the labels of the hexagonal lattices
to which each of these dimers belongs. If this sequence of labels is
$123123\cdots$, or a cyclic permutation thereof, we shall call the RGB loop
clockwise oriented; otherwise it is anticlockwise oriented.
These orientations can be used to define a one-dimensional height on
the lattice faces, which we shall refer to as the {\em RGB height}.
Properties of this height were studied in Ref.~\cite{Wilson02}.

Note however that the orientation constructed above is not an {\em intrinsic}
property of the RGB loops: unlike the orientation of $BC$-type loops
defined in section~\ref{sec_3C_model}, the RGB orientation is fixed once
the trajectory of the corresponding loop has been fixed. For this
reason we shall in the following rather focus on the orientations of
the $BC$-type loops, and on the associated height mapping.

Since the three copies of the fully-packed hexagonal-lattice O($n=1$) model do
not interact, the critical properties of RGB loops are easily found. The free
energies and the central charges of the three copies simply add up, and in
particular one has
\beq
 c = 3.
\label{RGB_c} 
\eeq

The same is true for critical exponents linked to vortex-type defects.
Correlation functions in the RGB model are products of those in the 
individual hexagonal-lattice models. So once a given defect configuration
on the triangular lattice has been decomposed into defects in the three
hexagonal-lattice models, the corresponding critical exponents just need to
be summed up.

\begin{figure}
\begin{center}
\epsfxsize=15.0cm \epsfysize=15.0cm
\epsfbox{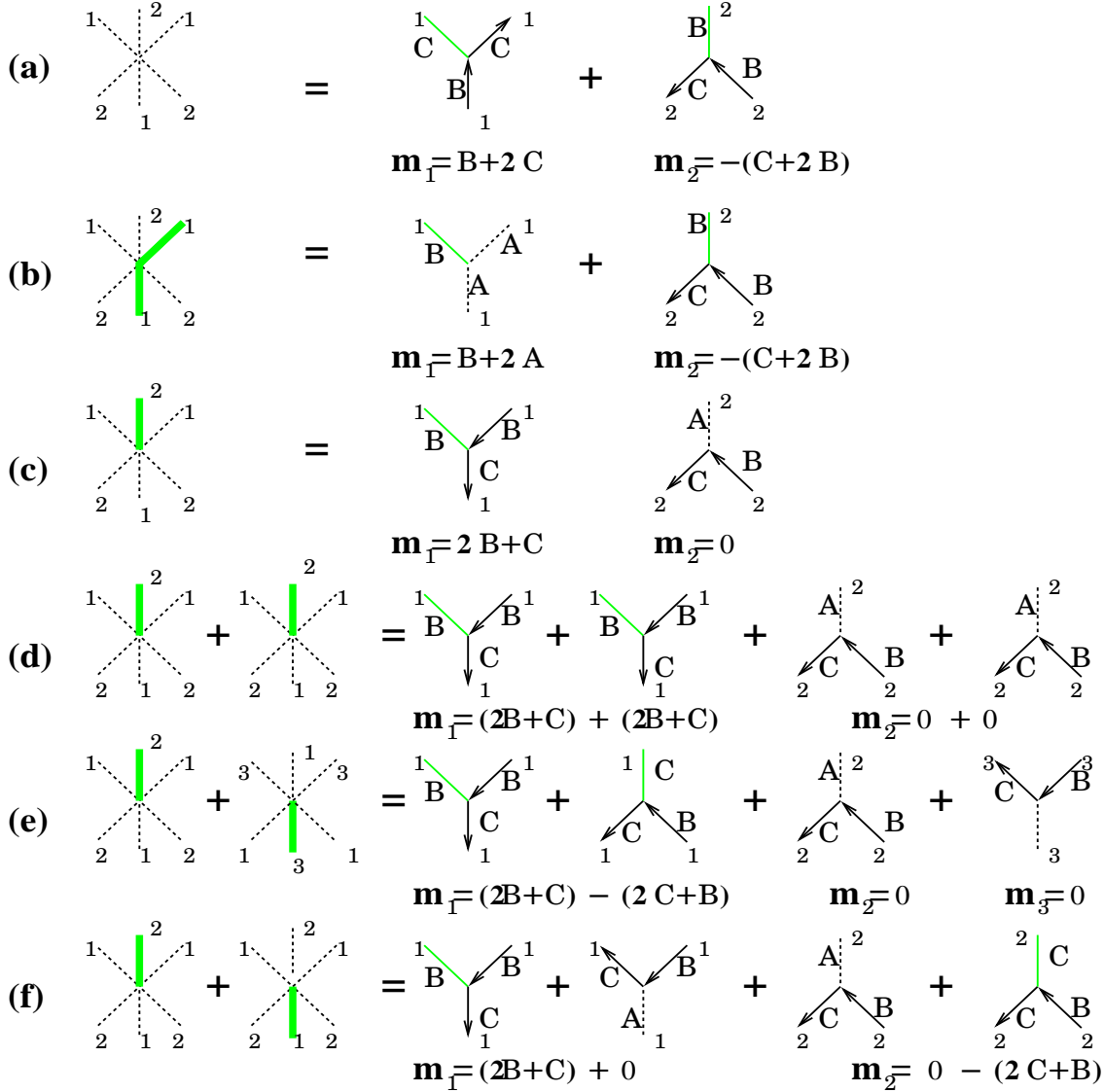}
\caption{Defect vortices on the triangular lattice (left) and on the
hexagonal lattice (right), with the corresponding magnetic charges.
$(a)$ Thermal defect, $(b)$ the $\pm{\pi/3}$ bend defect, $(c)$ one-string
defect, $(d)$--$(f)$ various examples of two-string defects (see text).
On the left, the RGB loops on the triangular lattice are shown in thick
grey linestyle, and the empty bonds by dotted lines.
On the right, the string defects on the hexagonal lattices (with labels
as shown) are given in thin grey linestyle; non-defect directed loops
are black. The magnetic
charges on the two sublattices are denoted by ${\bf m}_1$ and ${\bf m}_2$.}
\label{tridefects}
\end{center}
\end{figure}

In Fig.~\ref{tridefects} we show a few examples.
We first consider the thermal exponent on the triangular lattice, which
is associated with an empty vertex. The corresponding 
vortex configuration, shown in Fig.~\ref{tridefects}$(a)$, is equivalent
to two one-string defects on the two hexagonal lattices which meet at the
centre of the vortex. Thus, from Eq.~(\ref{string12}), 
\beq
 x^{\rm tri}_{T} = x^{\rm hex}_{1} + x^{\rm hex}_{1} = {1 \over 2}.
\label{RGB_xT}
\eeq  
Note that this is different from $x_T$ of the $n=1$ FPL models on the
hexagonal and the square lattices (which have $x_T=1$ and $x_T=2/3$
respectively). Also, in order for the energy-energy correlation function
to have zero total magnetic charge on {\em each} hexagonal lattice, the
two thermal defects have to placed on triangular-lattice vertices carrying
the same label. 

Next, consider the {\em bend defect} associated with a loop turning by an
otherwise forbidden angle of $\pm \pi/3$.
As all non-defect turns go through angles of $\pm 2\pi/3$, it is easy
to show that a loop cannot close if there is only one $\pm \pi/3$ defect.
We henceforth consider an RGB loop with two such defects, situated at distant
points, and corresponding to angles $+\pi/3$ and $-\pi/3$ of opposite sign.
The vortex due to one of these defects is
shown in Fig.~\ref{tridefects}$(b)$: it amounts to a one-string defect on
either of two hexagonal lattices, just as in the case of the thermal defect.
This leads to
\beq 
 x^{\rm tri}_{\pm \pi/3} = x^{\rm hex}_{1} + x^{h\rm ex}_{1} = {1 \over 2}.
\label{RGB_xbend}
\eeq 

The one-string defect on the triangular lattice is shown in
Fig.~\ref{tridefects}$(c)$. It is seen to be equivalent to a one-string
defect on one hexagonal lattice and no other defects on the other hexagonal
lattices. The critical exponent is therefore
\beq 
 x^{\rm tri}_{1} = x^{\rm hex}_{1} = {1 \over 4}.
\label{RGB_x1}
\eeq 

As discussed earlier, the fractal dimension $D_{\rm f}$ of a loop is such
that $2-D_{\rm f}$ (the codimension) is the critical exponent describing the
decay of the probability that two distant points belong to the same loop
\cite{SalDup87,Kondev00}. In a model of oriented loops, this critical
exponent can be computed by reversing the direction along one half of the
loop, whence it becomes identified with $x_2$ as in Eq.~(\ref{fractal}).
In the present case, the RGB loops do not possess an intrinsic orientation,
as discussed above,
and the two points must be marked in another way. A convenient way is
by inserting the bend defect of Fig.~\ref{tridefects}$(b)$. We therefore
arrive at
\beq
 D_{\rm f} = 2 - x^{\rm tri}_{\pm \pi/3} = \frac32.
\eeq
This agrees with earlier numerical work \cite{unpub}.
Note that this is different from $D_{\rm f}$ of the $n=1$ FPL models on the
hexagonal and the square lattices (which both have $D_{\rm f}=7/4$).

There are several different ways of making two strings emanate from
nearby, but distinct points. The corresponding critical exponents are
thus {\em not} linked to $D_{\rm f}$, and moreover depend on the exact
geometry of the string insertions, even though macroscopically the
two strings appear to have common end-points. We illustrate this comment
by a few examples.

Consider first the geometry of Fig.~\ref{tridefects}$(d)$ in which the two
nearby defects reside on the same sublattice and have the same alignment. The
defect charge is equivalent to that of two nearby one-string defects on one
hexagonal lattice, and leads to $x^{\rm tri}_{(d)} = x^{\rm hex}({\bf e}_{0},
2 (2{\bf B} + {\bf C})) = 5/4$, by Eq.~(\ref{dimension}).

A different alignment of the strings is shown in
Fig.~\ref{tridefects}$(e)$--$(f)$. If the two defects reside on
different sublattices, as in Fig.~\ref{tridefects}$(e)$, one obtains
$x^{\rm tri}_{(e)} = x^{\rm hex}({\bf e}_{0},({\bf B} - {\bf C})) = 1/4$.
If they are on the same sublattice, as in Fig.~\ref{tridefects}$(f)$, one has
$x^{\rm tri}_{(f)} = 1/2$.

\subsection{The triple RGB model}

A more exotic model can be obtained if we set $n=2$, so as to weigh equally
the three-colouring configurations (rather than the dimer coverings)
on each of the three hexagonal lattices. Each vertex on the triangular
lattice is now visited by three distinct RGB-type loops, having respective
colours $A$, $B$ and $C$.

A sample configuration is shown in Fig.~\ref{RGB}$(a)$, where the $A$-loops
appear as thick grey lines, the $B$-loops as thick black lines, and the
$C$-loops as thin black lines. The central charge of this model is now
$c= 3 \times 2 = 6$.

Again, we emphasise that the three copies of the $n=2$ hexagonal-lattice
FPL model are completely decoupled, and critical exponents can be obtained
along the lines laid out above.

\subsection{Staggered models}

Yet other models can be obtained by taking the $w\to\infty$ limit
in one or several of the superposed hexagonal-lattice FPL models.
As we have seen in section~\ref{sec_TM_w_inf} this will eliminate
one unit of central charge from each of the concerned models.

We recall that infinite $w$ means that the $E_0$-type edges cannot
have the colour $A$. Thus, within the RGB model one obtains critical
theories with $c=2$, $1$ or $0$, depending on whether the loops are
excluded from the $E_0$ edges on one, two or three hexagonal lattices.
Similarly, within the triple RGB model one gets $c=5$, $4$ or $3$,
depending on whether the $A$-coloured RGB loops are
excluded from the $E_0$ edges on one, two or three hexagonal lattices.
In all cases, critical exponents can be worked out as usual.


\section{Conclusion}
\label{sec_conclusion}

In this paper, we have studied the loop model (\ref{model}) on the hexagonal
lattice subjected to a staggered field $w$. The parameter $w$ breaks down
the three-phase coexistence present at $w=1$ and has an interesting
competition with the temperature $T$.

We have found that the model has four new critical branches for $w \neq 1$, as
seen in Fig.~\ref{phasedia}. Three of these exist for $0 \leq n \leq 2$ and
are denoted by {\sf P} (Potts model), {\sf D'} (tricritical dense loops) and
{\sf d'} (tricritical dilute loops). Furthermore, there is a fourth critical
point for $n > 2$ belonging to the Ising universality class. These are in
addition to the four old branches known to exist at $w=1$: {\sf C} (compact
loops), {\sf D} (dense loops) and {\sf d} (dilute loops) for $|n| \leq 2$; and
the critical three-state Potts model for $n > 2$.

It seems that the multicritical branches ${\sf D'}$ and ${\sf d'}$ are as
generic as the ${\sf D}$ and ${\sf d}$ branches, since all of these have
previously been found on the square and triangular lattice $O(n)$ models. An
explanation of this universality is given within the Coulomb gas formalism in
section~\ref{sec_CG_multi}. On the other hand, the point {\sf C} (compact
loops, $T=0$ and $w=1$) is specific to the hexagonal lattice. Due to the
enhanced symmetry, its Coulomb gas is based on a two-component height field,
and this is unstable towards both the perturbations in $T$ and in $w$.

A part of our results can be explained via the embedding of the model
(\ref{model}) into the more general triangular-lattice O($n$) model. However,
the classification of possible types of critical behaviour in this latter
model \cite{Nienhuis98} does not appear to be complete. In particular, compact
loops and the RGB model are interesting special cases of the
triangular-lattice O($n$) model which are not accounted for in
Ref.~\cite{Nienhuis98}.

It should be emphasised that even though the O($n$)-type loop models on
various lattices (hexagonal, square and triangular) share many universal
features, they also fail to account for some aspects of the low-temperature
physics of the original O($n$) spin model. For example, the generic
symmetry-broken low-temperature Goldstone phase is absent
\cite{ReadSal01,JRS03}, and for $n=2$ there is no line of critical points
below the Kosterlitz-Thouless temperature \cite{Blote89}.

Finally, we have also studied constrained compact RGB-type loops on the
triangular lattice. In particular, we have proved that RGB loops have the
fractal dimension $D_{\rm f}=3/2$. This finding would seem to indicate that
these loops are level lines of a Gaussian surface. More precisely, we
conjecture that the RGB height (defined in section~\ref{sec_RGB_model}) defines
a Gaussian surface.

Interestingly, our derivation of the value of $D_{\rm f}$ is based on level
lines of three different four-dimensional surfaces which are manifestly
non-Gaussian. It would be interesting if one could find a more direct
derivation, in terms of the RGB height. \\

\noindent
{\bf Acknowledgments}

It is a pleasure to thank Bernard Nienhuis and David Wilson
for some very helpful comments.

\end{document}